\global\def\draftcontrol{0}

   \def\versionno{Chemical Potential in Supersymmetric Field Theories in Curved Spaces}

\catcode`\@=11

\expandafter\ifx\csname draftcontrol\endcsname\relax\global\def\draftcontrol{0}
\fi

{\count255=\time\divide\count255 by 60
\xdef\hourmin{\number\count255}
\multiply\count255 by-60\advance\count255 by\time
\xdef\hourmin{\hourmin:\ifnum\count255<10 0\fi\the\count255}}
\def\draftdate{\number\month/\number\day/\number\year\ \ \ \hourmin }

\newcommand\makepapertitle{\par
  \begingroup
    \renewcommand\thefootnote{\@fnsymbol\c@footnote}%
    \def\@makefnmark{\rlap{\@textsuperscript{\normalfont\@thefnmark}}}%
    \long\def\@makefntext##1{\parindent 1em\noindent
            \hb@xt@1.8em{%
                \hss\@textsuperscript{\normalfont\@thefnmark}}##1}%
     \newpage
     \global\@topnum\z@   
     \@makepapertitle
     \thispagestyle{empty}\@thanks
  \endgroup
  \setcounter{footnote}{0}%
  \global\let\thanks\relax
  \global\let\makepapertitle\relax
  \global\let\@makepapertitle\relax
  \global\let\@thanks\@empty
  \global\let\@author\@empty
  \global\let\@date\@empty
  \global\let\@title\@empty
  \global\let\title\relax
  \global\let\author\relax
  \global\let\date\relax
  \global\let\and\relax
  \def\version{\let\version\@version\@gobble}
}
\def\@makepapertitle{%
  \newpage
   \ifnum\draftcontrol=1 {}
   \version\versionno
   \vskip 3em%
   \else
   \hfill\hbox to 3cm {\parbox{4cm}{\@pubnum}\hss}%
   \vskip 3em%
   \fi
   \begin{center}%
   \let \footnote \thanks
     {\LARGE {\@title}}%
     \vskip 1.5em%
     {\normalsize
       \lineskip .5em%
       \begin{tabular}[t]{c}%
         \@author
       \end{tabular}\par}%
     \vskip 1.5em%
     {\@bstract}%
     \end{center}%
     \vskip 1.5em
     \@date%
   \par
}

\gdef\@pubnum{}
\def\pubnum#1{%
  \gdef\@pubnum{#1}}

\gdef\@bstract{}
\def\Abstract#1{%
  \gdef\@bstract{%
   \parbox{\textwidth-0pc}{%
   \centerline{\bf Abstract}\penalty1000%
\noindent
\renewcommand\baselinestretch{1.0}%
{#1}}}
}

\def\ps@paper{\let\@mkboth\@gobbletwo%
     \ifnum\draftcontrol=1
        \def\@oddfoot{\hbox to \textwidth{\tiny \versionno \hfil\tiny\draftdate}%
        \hskip -\textwidth \hbox to \textwidth{\hfil\rm\thepage\hfil}}%
     \else\def\@oddfoot{\hbox to \textwidth{\hfil\rm\thepage\hfil}}
     \fi
     \let\@evenfoot\@oddfoot
}

\def\@version#1{\ifnum\draftcontrol=1
\typeout{}\typeout{#1}\typeout{}
\vskip3mm\centerline{\hbox{\fbox{\normalsize{\tt DRAFT -- #1 -- }
                   {\draftdate}}}}\vskip3mm
\fi}
\let\version\@version
\long\def\eqlabel#1{\ifnum\draftcontrol=1
                    \tag@false  
                    \tag*{(\theequation) \hbox to -0.2cm{\hspace{0cm}\small{#1}\hss}}
                    \refstepcounter{equation}
                    \edef\@currentlabel{\theequation}
                    \ltx@label{#1}          
                    \else
                    \label{#1}
                    \fi
                    }
\let\st@bibitem\@bibitem
\let\st@lbibitem\@lbibitem
\ifnum\draftcontrol=1
  \def\@bibitem#1{%
    \st@bibitem{#1}\a@@label{#1}\ignorespaces}
  \def\@lbibitem[#1]#2{%
    \st@lbibitem[#1]{#2}\a@@label{#2}\ignorespaces}
  \def\a@@label#1{%
    \gdef\a@lab{\smash{\normalfont\small#1}}
    \ifvmode
      \if@inlabel
        \global\setbox\@labels\hbox{%
          \llap{\a@lab\let\a@lab\relax
                \kern\@totalleftmargin\kern\marginparsep}%
          \box\@labels}%
      \fi
    \fi}
\fi

\documentclass[12pt,letterpaper]{article}

\usepackage{amssymb,array,calc,rotating,epsfig,psfrag}
\usepackage[nosort]{cite}
\usepackage{bbold}
\usepackage{geometry}

\usepackage[centertags]{amsmath}
\usepackage{amsfonts}
\usepackage{amscd,amsthm}
\usepackage{amssymb}
\usepackage{geometry}

\ifnum\draftcontrol=1
\fi

\renewcommand\baselinestretch{1.25}
\renewcommand\section{\@startsection {section}{1}{\z@}%
                                   {-3.5ex \@plus -1ex \@minus -.2ex}%
                                   {2.3ex \@plus.2ex}%
                                   {\normalfont\large\bfseries}}
\renewcommand\subsection{\@startsection{subsection}{2}{\z@}%
                                   {-3.25ex\@plus -1ex \@minus -.2ex}%
                                   {1.5ex \@plus .2ex}%
                                   {\normalfont\normalsize\bfseries}}
\renewcommand\subsubsection{\@startsection{subsubsection}{3}{\z@}%
                                   {-3.25ex\@plus -1ex \@minus -.2ex}%
                                   {1.5ex \@plus .2ex}%
                                   {\normalfont\normalsize\it}}
\renewcommand\paragraph{\@startsection{paragraph}{4}{\z@}%
                                   {-3.25ex\@plus -1ex \@minus -.2ex}%
                                   {1.5ex \@plus .2ex}%
                                   {\normalfont\normalsize\bf}}

\def\revise#1       {\raisebox{-0em}{\rule{3pt}{1em}}%
                     \marginpar{\raisebox{.5em}{\vrule width3pt\
                     \vrule width0pt height 0pt depth0.5em
                     \hbox to 0cm{\hspace{0cm}{%
                     \parbox[t]{4em}{\raggedright\footnotesize{#1}}}\hss}}}}

\def\del          {\partial}

\def\tr           {\mathop{\rm Tr}}
\def\Re           {{\rm Re\hskip0.1em}}
\def\Im           {{\rm Im\hskip0.1em}}

\def\half{{\frac12}}

\def\sqr#1#2{{\vcenter{\vbox{\hrule height.#2pt
 \hbox{\vrule width.#2pt height#1pt \kern#1pt
 \vrule width.#2pt}\hrule height.#2pt}}}}

\newcommand{\fft}[2]{{\frac{#1}{#2}}}
\newcommand{\ft}[2]{{\textstyle{\frac{#1}{#2}}}}

\def\a{\alpha}
\def\b{\beta}

\def\r{\rho}

\def\o{\omega}

\def\m{\mu}
\def\g{\gamma}
\def\l{\lambda}
\def\n{\nu}
\def\bn{\bar{\nu}}
\def\bm{\bar{\mu}}

\usepackage{color}

\catcode`\@=12

\begin{document}



\newcommand{\be}{\begin{equation}}
\newcommand{\ee}{\end{equation}}
\newcommand{\beq}{\begin{equation}}
\newcommand{\eeq}{\end{equation}}
\newcommand{\ba}{\begin{eqnarray}}
\newcommand{\ea}{\end{eqnarray}}
\newcommand{\nn}{\nonumber}

\def\vol{\bf vol}
\def\Vol{\bf Vol}
\def\del{{\partial}}
\def\vev#1{\left\langle #1 \right\rangle}
\def\cn{{\cal N}}
\def\co{{\cal O}}
\def\IC{{\mathbb C}}
\def\IR{{\mathbb R}}
\def\IZ{{\mathbb Z}}
\def\RP{{\bf RP}}
\def\CP{{\bf CP}}
\def\Poincare{{Poincar\'e }}
\def\tr{{\rm tr}}
\def\tp{{\tilde \Phi}}
\def\Y{{\bf Y}}
\def\te{\theta}
\def\bX{\bf{X}}

\def\TL{\hfil$\displaystyle{##}$}
\def\TR{$\displaystyle{{}##}$\hfil}
\def\TC{\hfil$\displaystyle{##}$\hfil}
\def\TT{\hbox{##}}
\def\HLINE{\noalign{\vskip1\jot}\hline\noalign{\vskip1\jot}} 
\def\seqalign#1#2{\vcenter{\openup1\jot
  \halign{\strut #1\cr #2 \cr}}}
\def\lbldef#1#2{\expandafter\gdef\csname #1\endcsname {#2}}
\def\eqn#1#2{\lbldef{#1}{(\ref{#1})}%
\begin{equation} #2 \label{#1} \end{equation}}
\def\eqalign#1{\vcenter{\openup1\jot
    \halign{\strut\span\TL & \span\TR\cr #1 \cr
   }}}
\def\eno#1{(\ref{#1})}
\def\href#1#2{#2}
\def\half{{1 \over 2}}

\def\ads{{\it AdS}}
\def\adsp{{\it AdS}$_{p+2}$}
\def\cft{{\it CFT}}

\newcommand{\ber}{\begin{eqnarray}}
\newcommand{\eer}{\end{eqnarray}}

\newcommand{\bea}{\begin{eqnarray}}
\newcommand{\eea}{\end{eqnarray}}

\newcommand{\beqar}{\begin{eqnarray}}
\newcommand{\cN}{{\cal N}}
\newcommand{\cO}{{\cal O}}
\newcommand{\cA}{{\cal A}}
\newcommand{\cT}{{\cal T}}
\newcommand{\cC}{{\cal C}}
\newcommand{\cR}{{\cal R}}
\newcommand{\cW}{{\cal W}}
\newcommand{\eeqar}{\end{eqnarray}}
\newcommand{\lm}{\lambda}\newcommand{\Lm}{\Lambda}
\newcommand{\eps}{\epsilon}


\newcommand{\nonu}{\nonumber}
\newcommand{\oh}{\displaystyle{\frac{1}{2}}}
\newcommand{\dsl}
  {\kern.06em\hbox{\raise.15ex\hbox{$/$}\kern-.56em\hbox{$\partial$}}}
\newcommand{\as}{\not\!\! A}
\newcommand{\ps}{\not\! p}
\newcommand{\ks}{\not\! k}
\newcommand{\D}{{\cal{D}}}
\newcommand{\dv}{d^2x}
\newcommand{\Z}{{\cal Z}}
\newcommand{\N}{{\cal N}}
\newcommand{\Dsl}{\not\!\! D}
\newcommand{\Bsl}{\not\!\! B}
\newcommand{\Psl}{\not\!\! P}
\newcommand{\eeqarr}{\end{eqnarray}}
\newcommand{\ZZ}{{\rm \kern 0.275em Z \kern -0.92em Z}\;}

\def\s{\sigma}
\def\a{\alpha}
\def\b{\beta}
\def\r{\rho}
\def\d{\delta}
\def\g{\gamma}
\def\G{\Gamma}
\def\ep{\epsilon}
\makeatletter \@addtoreset{equation}{section} \makeatother
\renewcommand{\theequation}{\thesection.\arabic{equation}}

\def\be{\begin{equation}}
\def\ee{\end{equation}}
\def\bea{\begin{eqnarray}}
\def\eea{\end{eqnarray}}
\def\m{\mu}
\def\n{\nu}
\def\g{\gamma}
\def\p{\phi}
\def\L{\Lambda}
\def \W{{\cal W}}
\def\bn{\bar{\nu}}
\def\bm{\bar{\mu}}
\def\bw{\bar{w}}
\def\ba{\bar{\alpha}}
\def\bb{\bar{\beta}}

\newcommand{\tbyo}[2]{\begin{pmatrix}#1\\#2 \end{pmatrix}}
\newcommand{\obyt}[2]{\begin{pmatrix}#1 & #2 \end{pmatrix}}
\newcommand{\tbyt}[4]{\begin{pmatrix}#1 & #2 \\ #3 & #4 \end{pmatrix}}
\newcommand{\pd}{\partial}
\newcommand{\pdf}[2]{\frac{\partial#1}{\partial#2}}
\newcommand{\fdf}[2]{\frac{\delta #1}{\delta #2}}
\newcommand{\idn}{{1\relax{\kern-.35em}1}}


\renewcommand\d{\delta}
\newcommand\e{\varepsilon}
\newcommand\wt{\widetilde}
\newcommand{\abs}[1]{\left\vert#1\right\vert}
\newcommand\cH{\mathcal{H}}
\newcommand\cF{\mathcal{F}}
\newcommand\cB{\mathcal{B}}
\newcommand{\w}{\omega}
\newcommand{\slh}[1]{#1\relax{\kern-.9em}\slash}

\begin{titlepage}

\version\versionno

\rightline{\small{\tt MCTP-12-16}}

\vskip 1.7 cm

\centerline{\bf \Large  Rigid Supersymmetric Backgrounds  }

\vskip .5cm

\centerline{\bf \Large of Minimal Off-Shell Supergravity}

\vskip 1cm

\centerline{\large James T. Liu, Leopoldo A. Pando Zayas and Dori Reichmann}

\vskip .8cm
\centerline{\it Michigan Center for Theoretical
Physics}
\centerline{ \it Randall Laboratory of Physics, The University of
Michigan}
\centerline{\it Ann Arbor, MI 48109-1040}

\vspace{1cm}

\begin{abstract}
We discuss various aspects of rigid supersymmetry within minimal  ${\cal N} = 1$
off-shell supergravity using the old and new minimal
formulations both in Lorentzian and Euclidean signatures. In particular, we
construct all rigid supersymmetry backgrounds with a hypersurface
orthogonal Killing vector.
In the Lorentzian signature we show that $AdS_4$ provides a rigid
supersymmetric background in both formulations albeit with different
amounts of preserved supersymmetry.
In the Euclidean signature we find new backgrounds of the old-minimal
supergravity, including squashed four-spheres and a half-BPS version of flat
space.
\end{abstract}



\end{titlepage}




\section{Introduction}
Recently, two important problems were solved by considering supersymmetric field theories on compact spaces.  The first is the use of localization techniques by Pestun to compute the exact expectation value of the half-supersymmetric circular Wilson loop in ${\cal N}=4$ supersymmetric Yang-Mills  \cite{Pestun:2007rz}. This result provided a rigorous proof of the conjecture stating that such expectation values can be computed using a Gaussian matrix model \cite{Drukker:2000rr,Erickson:2000af}. The second was the insight gained into the number of degrees of freedom of some three-dimensional field theories, i.e.\ the $N^{3/2}$ problem. The breakthrough was achieved, again, by considering supersymmetric field theories on $S^3$  \cite{Kapustin:2009kz}  and subsequently studying the free energy in the resulting matrix model \cite{Drukker:2010nc}.

One particularly interesting development motivating our work is the study of supersymmetric field theories on squashed spheres. Namely, Ref.~\cite{Hama:2011ea} studied Euclidean 3D ${\cal N}=2$ supersymmetric gauge theories on squashed three-spheres, computed the partition function using localization and found a precise dependence on the squashing parameter. A similar calculation was performed in \cite{Imamura:2011wg}, and an explicit calculation of the large $N$ limit yielded a free energy  scaling as $N^{3/2}$. The gravity duals of \cite{Hama:2011ea}  and \cite{Imamura:2011wg} were presented in \cite{Martelli:2011fu} and \cite{Martelli:2011fw} respectively  and exact agreement was found for the computation of the free energy via comparison with the gravity free energy.

These developments naturally raise the general question of how to describe supersymmetric field theories in a fixed (rigid) curved background.
A uniform treatment of rigid supersymmetric field theories in curved space was initiated by Festuccia and Seiberg in \cite{Festuccia:2011ws}. The main idea is
to start with an off-shell supergravity theory, then decouple gravity in order to obtain a theory that is rigidly supersymmetric.  Demanding that the supersymmetry variation of the gravitino vanishes constrains the possible spacetime manifolds where a supersymmetric field theory can be defined. In this approach, the auxiliary fields of the supergravity multiplet must be chosen appropriately to solve the gravitino variation. Such a choice then determines the couplings in the resulting rigid field theory. The discussion of rigid supersymmetry initiated in \cite{Festuccia:2011ws} has been extended to various situations in subsequent work
\cite{Jia:2011hw,Samtleben:2012gy,Dumitrescu:2012ha,Klare:2012gn,Cassani:2012new}.

While the localization techniques are unique to compact Riemannian spaces, another interesting point of view comes from considering the Festuccia-Seiberg construction in Lorentzian signature. The field theories with rigid partial supersymmetry give an interesting window to the study supersymmetric field theories with broken Poincar\'e symmetry. These have possible application to the study of holographic duals away from the vacuum solution.

As discussed in \cite{Festuccia:2011ws}, in four dimensions there are two formulations for off-shell supergravity that lead to different sets of auxiliary fields and Killing spinor equations. In the old-minimal formulation the auxiliary fields are a scalar $S$, a pseudoscalar $P$ and a vector $V$,  giving the Killing spinor equation
\begin{equation}
[\nabla_\mu-\ft{i}6(\gamma_\mu{}^\nu-2\delta_\mu^\nu)
\gamma_5V_\nu+\ft16\gamma_\mu(S+i\gamma_5P)]\epsilon = 0.
\end{equation}
For the new-minimal formulation the auxiliary fields are a gauge field $A$ and a vector $V$, giving the Killing spinor equation
\begin{equation}
[\nabla_\mu-\ft{i}2(\gamma_\mu{}^\nu-2\delta_\mu^\nu)\gamma^5V_\nu
+i\gamma_5A_\nu]\epsilon = 0.
\end{equation}
In this paper we explore various aspects of these equations, specifically clarifying the difference between the Euclidean and Lorentzian signatures. We adopt the spinor bilinear techniques of \cite{Tod:1983pm,Gauntlett:2002nw} as applied in this context by the authors of \cite{Samtleben:2012gy, Dumitrescu:2012ha,Klare:2012gn}. In particular, we find that if the
background admits a Killing vector $K$, it necessarily obeys the additional differential constraint
\begin{equation}
    K\wedge dK \propto *K(i_KV).
\end{equation}
In the cases where $i_KV=0$, the left-hand side implies that the Killing vector is hypersurface orthogonal. In these cases we can construct explicit forms for all possible solutions. Among these solutions we find maximal symmetric spaces $\mathbb{R}^4$, $\mathrm{AdS}_4$, $\mathbb{H}^4$ and $\mathrm{S}^4$ and some of their partial supersymmetric deformations.

The paper is organized as follows. In section \ref{Sec:OldMin} we discuss the old minimal off-shell supergravity in both its Lorentzian and Euclidean formulations. We show that the Lorentzian formulation admits $AdS_4$ as  a rigid supersymmetric background and that the Euclidean formulation admits the squashed $\mathrm{S}^4$.
In section  \ref{Sec:NewMin} we discuss the new minimal off-shell supergravity. Given that the auxiliary fields are two vectors, a natural expectation is that the maximally symmetric space $AdS_4$ would not be a solution to the rigid supersymmetry conditions. We show that, contrary to this intuition, $AdS_4$ is actually a supersymmetric background of the new minimal supergravity albeit with less preserved supersymmetry. For the new minimal formulation of supergravity in Euclidean signature, the question of rigid backgrounds has recently been  discussed in great detail in \cite{Dumitrescu:2012ha}  and \cite{Klare:2012gn}. We therefore focus on the Lorentzian signature analysis.
We present some concluding remarks in section \ref{Sec:Conclusions}. In a series of appendices we discuss various toy gravitino variations  that should be useful in understanding squashed sphere backgrounds in various dimensions.

\section{All supersymmetric backgrounds of old minimal supergravity}\label{Sec:OldMin}

The fields of old minimal supergravity \cite{Stelle:1978ye,Ferrara:1978em} consist of the graviton
$g_{\mu\nu}$ and gravitino $\psi_\mu$, along with the auxiliary fields $S$, $P$ and $V_\mu$ (scalar, pseudoscalar and axial vector, respectively).
It is well known that fermions, and hence the supersymmetry algebra, is sensitive to both
spacetime dimension and signature.  The standard formulation of supergravity is in a Lorentzian
spacetime, and we begin with this case.

\subsection{Lorentzian signature}

In $3+1$ dimensions, we may take the minimal spinor to be Majorana.  For a metric
with signature $(-,+,+,+)$, we may take a Majorana representation for the Dirac matrices,
so that they are real.  (Note, however, that $\gamma_5\equiv (i/4!)\epsilon_{\mu\nu\rho\sigma}
\gamma^\mu\gamma^\nu\gamma^\rho\gamma^\sigma$ is imaginary.)  In particular, $\gamma^0$
is real antisymmetric while $\gamma^i$ are real symmetric.  Hermitian conjugation and transposition take the form
\begin{equation}
\gamma_0\gamma_\mu\gamma^0=\gamma_\mu^\dagger,\qquad
C\gamma_\mu C^{-1}=-\gamma_\mu^t.
\end{equation}
In the Majorana representation, the charge conjugation matrix is simply
$C=\gamma^0$.  Recalling that the Dirac conjugate of $\psi$ is $\bar\psi=\psi^\dagger\gamma^0$
and that the Majorana conjugate is  $\psi^c=\psi^tC$, we see that the Majorana condition
$\psi^c=\bar\psi$ is satisfied for four-real component spinors $\psi=\psi^*$.

As a result, old minimal supergravity in a Lorentzian signature admits four real supercharges.
In the off-shell formulation, the gravitino has 16 real fermionic degrees of freedom, while the
bosonic fields have $10+1+1+4=16$ real bosonic degrees of freedom (for $g_{\mu\nu}$,
$S$, $P$ and $V_\mu$, respectively).  Up to a rescaling of fields, the gravitino variation is
given by
\begin{equation}
\delta\psi_\mu=\mathcal D_\mu\epsilon\equiv[\nabla_\mu-\ft{i}6(\gamma_\mu{}^\nu-2\delta_\mu^\nu)
\gamma_5V_\nu+\ft16\gamma_\mu(S+i\gamma_5P)]\epsilon.
\label{eq:deltapsi}
\end{equation}
Note that this expression is in fact real since $\gamma_5$ is imaginary.  We now wish to obtain
supersymmetric backgrounds in the context of old minimal supergravity.  What this means is that
we would like to find a background metric $g_{\mu\nu}$ along with the scalar, pseudoscalar and
axialvector auxiliary fields $S$, $P$ and $V_\mu$ admitting (at least) one solution to the Killing
spinor equation, $\mathcal D_\mu\epsilon=0$.

Killing spinors $\epsilon$ must solve the integrability condition\footnote{We discuss various general aspects of integrability in appendix \ref{App:Integrability}.} $[\mathcal D_\mu,\mathcal D_\nu]
\epsilon=0$ where
\begin{eqnarray}
[\mathcal D_\mu,\mathcal D_\nu]\!\!&=&\!\!\ft14[R_{\mu\nu}{}^{\rho\sigma}
+\ft29(\delta_\mu^\rho\delta_\nu^\sigma V^2
-2\delta_{[\mu}^\rho V_{\nu]}V^\sigma)+\ft29\delta_\mu^\rho\delta_\nu^\sigma(S^2+P^2)
+\ft23\epsilon_{\alpha\beta}{}^{\rho\sigma}\delta_{[\mu}^\alpha\nabla_{\nu]}V^\beta]
\gamma_{\rho\sigma}\nonumber\\
&&-\ft13[\delta_{[\mu}^\rho(\partial_{\nu]}S+\ft23V_{\nu]}P)+\ft13\epsilon_{\mu\nu}{}^{\lambda\rho}
V_\lambda S]\gamma_\rho\nonumber\\
&&-\ft{i}3[\delta_{[\mu}^\rho(\partial_{\nu]}P-\ft23V_{\nu]}S)+\ft13\epsilon_{\mu\nu}{}^{\lambda\rho}
V_\lambda P]\gamma_\rho\gamma_5
+\ft{2i}3\partial_{[\mu}V_{\nu]}\gamma_5.
\label{eq:integr}
\end{eqnarray}
For completely unbroken supersymmetry, each of the quantities multiplying the different
Dirac matrix combinations in (\ref{eq:integr}) must vanish independently.  This gives rise to the
following conditions for completely unbroken supersymmetry recently discussed in \cite{Festuccia:2011ws}
\begin{eqnarray}
&&\nabla_\mu V_\nu=0,\qquad V_\mu S=V_\mu P=0,\qquad \partial_\mu S=\partial_\mu P=0,
\nonumber\\
&&R_{\mu\nu}=\ft29(V_\mu V_\nu-g_{\mu\nu}V^2)-\ft13g_{\mu\nu}(S^2+P^2),\nonumber\\
&&C_{\mu\nu\rho\sigma}=0.
\label{eq:unbroken}
\end{eqnarray}
In particular, the restrictions $V_\mu S=V_\mu P=0$ are highly constraining, and lead to two
classes of solutions, the first with $V_\mu=0$ and non-vanishing constant $S$ or $P$
and the second with $S=P=0$ and a covariantly constant vector \cite{Festuccia:2011ws}.
In the first case, we have
\begin{equation}
R_{\mu\nu}=-\ft13g_{\mu\nu}(S^2+P^2),\qquad C_{\mu\nu\rho\sigma}=0,
\end{equation}
and hence the background is AdS$_4$ with radius $3/\sqrt{S^2+P^2}$.

In general, all we really demand is at least one unbroken supersymmetry.  In this case, the
conditions (\ref{eq:unbroken}) are in general too restrictive.
We thus proceed with an invariant tensor analysis.  For a real (commuting) Killing
spinor $\epsilon$, we may form the bilinears
\begin{equation}
K_\mu=\bar\epsilon\gamma_\mu\epsilon,\qquad J_{\mu\nu}=\bar\epsilon\gamma_{\mu\nu}\epsilon.
\end{equation}
Since $\epsilon\otimes\bar\epsilon$ is a real symmetric $4\times4$ matrix, there are 10 distinct
bilinear components.  Hence $K_\mu$ (with four components) and $J_{\mu\nu}$ (with six components)
exhaust the bilinears.  Of course, these components are not all independent.  The important
algebraic identities between $K_\mu$ and $J_{\mu\nu}$ are
\begin{equation}
K_\mu K^\mu=0,\qquad J_\mu{}^\nu J_{\nu\rho}=-K_\mu K_\rho,\qquad i_KJ=0,
\qquad K\wedge J=0,\qquad
\label{eq:algident}
\end{equation}
Note that this allows us to write
\begin{equation}
J=K\wedge X,\qquad\mbox{where}\qquad X_\mu X^\mu=1,\qquad i_KX=0.
\label{eq:Xdef}
\end{equation}
Here $X^\mu$ is an auxiliary unit-norm spacelike vector.

The differential identities following from $\mathcal D_\mu\epsilon=0$ are
\begin{eqnarray}
\nabla_\mu K_\nu\!&=&\!-\ft13\epsilon_{\mu\nu}{}^{\rho\sigma}V_\rho K_\sigma
+\ft13(J_{\mu\nu}S+*J_{\mu\nu}P),\nonumber\\
\nabla_\mu J_{\nu\lambda}\!&=&\!\ft23(-V_\mu*J_{\nu\lambda}
+*J_{\mu[\nu}V_{\lambda]}-g_{\mu[\nu}*J_{\lambda]\alpha}V^\alpha)
+\ft23g_{\mu[\nu}K_{\lambda]}S-\ft13\epsilon_{\mu\nu\lambda\sigma}K^\sigma P.
\nonumber\\
\label{eq:diffident}
\end{eqnarray}
By symmetrizing and antisymmetrizing the first identity, we see that
\begin{equation}
\nabla_{(\mu}K_{\nu)}=0,\qquad dK=\ft23(i_K*V+JS+*JP).
\label{eq:dK}
\end{equation}
Along with (\ref{eq:algident}), this demonstrates that $K^\mu$ is a null Killing vector.  Note,
however, that it is not necessarily hypersurface orthogonal, as
\begin{equation}
K\wedge dK=-\ft23*K(i_KV).
\end{equation}
Before proceeding, we also note the identities
\begin{equation}
dJ=-*KP,\qquad d*J=*KS,
\label{eq:dJdsJ}
\end{equation}
that follow directly from the second identity of (\ref{eq:diffident}).  Of course, it is worth keeping in
mind that the covariant derivative of $J$ encodes additional information beyond that given in
(\ref{eq:dJdsJ}).

We now restrict to the case $i_KV=0$, so that $K$ is hypersurface orthogonal.  It is also easy to
see that $K^\mu\nabla_\mu K_\nu=0$ holds (even in general).  This allows us to introduce
specialized coordinates $(u,v,y^m)$ and a function $H(u,y^m)$ so that
\begin{equation}
K^\mu\partial_\mu=\partial_v,\qquad K_\mu dx^\mu=H^{-1}du.
\label{eq:Kprop}
\end{equation}
Without loss of generality, we now write the metric in the form
\begin{equation}
ds^2=H^{-1}[\mathcal Fdu^2+2du\,dv+\hat g_{mn}dy^mdy^n],
\label{eq:metans}
\end{equation}
and take a vierbein basis
\begin{equation}
e^+=H^{-1}du,\qquad e^-=dv+\ft12\mathcal Fdu,\qquad e^a=H^{-1/2}\hat e_m^ady^m.
\end{equation}
In this case, the invariant tensors take the form
\begin{eqnarray}
&&K=e^+,\kern4.8em *K=e^+\wedge e^1\wedge e^2,\nonumber\\
&&J=X_ae^+\wedge e^a,\qquad *J=X_a\epsilon_{ab}e^+\wedge e^b,
\end{eqnarray}
where $\epsilon_{+-12}=1$, and where $X_a$ is introduced following (\ref{eq:Xdef}) and
satisfies $X_aX_a=1$.  For the auxiliary fields, no expansion is needed for $S$ and $P$.  For
$V_\mu$, we write
\begin{equation}
V=V_+e^++V_ae^a,\qquad *V=V_+e^+\wedge e^1\wedge e^2
-V_a\epsilon_{ab}e^+\wedge e^-\wedge e^b.
\end{equation}
We are now tasked with finding the conditions on the metric fields $H$, $\mathcal F$ and
$\hat g_{mn}$, the auxiliary fields $S$, $P$ and $V$ and the spacelike unit-norm vector $X_a$
such that the solution is supersymmetric.

Instead of directly solving the Killing spinor equation $\mathcal D_\mu\epsilon=0$, we make
use of the differential identities (\ref{eq:diffident}).  Starting with $K=e^+$, we find
\begin{equation}
dK=H^{-1/2}\partial_mH\hat e^m_ae^+\wedge e^a.
\end{equation}
Using (\ref{eq:dK}), and decomposing along $X_a$ and orthogonal to $X_a$ (which is
well defined in the two-dimensional transverse space), we find
\begin{eqnarray}
S&=&\ft32H^{-1/2}X^m\partial_mH+\hat\epsilon^{mn}X_mV_n,\nonumber\\
P&=&\ft32H^{-1/2}\hat\epsilon^{mn}X_m\partial_nH-X^mV_m.
\label{eq:SPfromK}
\end{eqnarray}
Here the curved-space indices on $X_m$ and $V_m$ are obtained using the zweibein $\hat e_m^a$.
Turning to $dJ$ and $d*J$ in (\ref{eq:dJdsJ}), we find
\begin{equation}
S=-H^2\hat\nabla^m(H^{-3/2}X_m),\qquad P=H^2\hat\epsilon^{mn}\partial_m(H^{-3/2}X_n).
\label{eq:SPsol}
\end{equation}
Combining this with (\ref{eq:SPfromK}) then allows us to extract
\begin{equation}
V_m=H^{1/2}[-X_m(\hat\epsilon^{np}\partial_nX_p)+\hat\epsilon_{mn}X^n(\hat\nabla^pX_p)].
\label{eq:Amsol}
\end{equation}

We have now solved for all auxiliary field components except for $V_+$.  To obtain $V_+$, we
must turn to the differential identity for $\nabla_+J_{+a}$.  This allows us to determine
\begin{equation}
V_+=-H\hat\epsilon_{mn}X^m\partial_uX^n.
\label{eq:Apsol}
\end{equation}
This has now exhausted the information contained in the differential identities (\ref{eq:diffident}).
As a result, we have constructed a supersymmetric background starting with a metric of the form
(\ref{eq:metans}), with three arbitrary functions $H(u,y^m)$, $\mathcal F(u,y^m)$ and
$\hat g_{mn}(u,y^m)$.  In addition to these three functions, the solution is parameterized by a
unit spacelike vector $X_m(u,y^m)$ satisfying $X_mX^m=1$.  At first glance, it is perhaps
rather surprising that the metric is completely arbitrary, other than that it admits a null Killing vector
satisfying (\ref{eq:Kprop}).  However, recall that in the off-shell formulation there is no need to
make use of any equations of motion.  For a given background specified by the metric and
$X_m$, the auxiliary fields are given by (\ref{eq:SPsol}), (\ref{eq:Amsol}) and (\ref{eq:Apsol}),
and no further conditions are needed for supersymmetry.  Of course, once the equations of
motion are given, then they will further restrict the backgrounds beyond the supersymmetry
analysis presented here.

In order to obtain the Killing spinor $\epsilon$, we note that a Fierz rearrangement allows us
to show that
\begin{equation}
K_\mu\gamma^\mu\epsilon=0.
\end{equation}
Since we have taken $K_+=1$, this corresponds to a projection $\gamma^+\epsilon=0$,
or equivalently $\gamma_-\epsilon=0$.  This projection preserves half of the original four
supersymmetries.  However, examination of the Killing spinor equation following from
(\ref{eq:deltapsi}) demonstrates that a second projection $X_a\gamma^a\epsilon=\epsilon$
is required as well.  The simultaneous conditions
\begin{equation}
\gamma^+\epsilon=0,\qquad X_a\gamma^a\epsilon=\epsilon,
\label{eq:BPSproj}
\end{equation}
demonstrate that this background generically preserves one of the four supersymmetries.
(In some cases, where the symmetry is enhanced, the background will admit more Killing
spinors than just the one we have {\it a priori} postulated.)

The two-dimensional metric $\hat g_{mn}$ can always be chosen to be conformally
flat.  Fixing the gauge $\hat g_{mn}=e^{2\sigma}\delta_{mn}$ simplifies some of the expressions
for the auxiliary fields
\begin{eqnarray}
S&=&-H^2e^{-2\sigma}\partial_a(H^{-3/2}e^\sigma X_a), \qquad P=H^2e^{-2\sigma}\epsilon_{ab}\partial_a(H^{-3/2}e^\sigma X_b),\nonumber\\
V_+&=&-H\epsilon_{ab}X_a\partial_uX_b, \qquad V_m=H^{1/2}[-\epsilon_{ab}X_a\partial_mX_b+\hat\epsilon_m{}^n\partial_n\sigma].
\label{eq:cgaux}
\end{eqnarray}
In addition, this choice allows us to obtain the explicit Killing spinor
\begin{equation}
\epsilon=e^{-\fft\alpha2\gamma_{12}}\epsilon_0,\qquad\gamma^1\epsilon_0=\epsilon_0,
\qquad\gamma^+\epsilon_0=0,
\label{eq:explicitks}
\end{equation}
where $\tan\alpha=X_2/X_1$, and where $\epsilon_0$ is a constant spinor satisfying the
projections (\ref{eq:BPSproj}).

\subsubsection{Solution with a constant Killing spinor}

Since $X_a$ is a unit vector in $\mathbb R^2$, it takes values in U(1), where $\alpha$
given above is the phase.  In general, $\alpha$ may depend on $y^1$, $y^2$ and $u$.
However, as a special case, we may consider solutions with constant $\alpha$.  This may
be obtained by setting, {\it e.g.}, $X_1=1$ and $X_2=0$.  The expressions for the auxiliary
fields, (\ref{eq:cgaux}), then reduce to
\begin{eqnarray}
S&=&-H^2e^{-2\sigma}\partial_1(H^{-3/2}e^\sigma),\qquad P=-H^2e^{-2\sigma}\partial_2(H^{-3/2}e^\sigma),\nonumber\\
V_+&=&0,\qquad V_m=H^{1/2}\hat\epsilon_m{}^n\partial_n\sigma.
\end{eqnarray}
In this case, the vector field $V_m$ lies entirely in the two-dimensional space spanned by
$(y^1,y^2)$, and points along equal $\sigma$ contours.

Note that the vector field vanishes for constant $\sigma$ or when $\sigma(u)$ depends at
most only on $u$.  This leads to a further simplification
\begin{equation}
S=3e^{-\sigma(u)}\partial_1\sqrt{H(u,y^1,y^2)},\qquad
P=3e^{-\sigma(u)}\partial_2\sqrt{H(u,y^1,y^2)},\qquad V_\mu=0.
\end{equation}
The maximally symmetric AdS$_4$ background fits into this case.  In particular, if we take
$\sigma=0$ and $H=(y^1/L)^2$, where $L$ is a constant, we end up with the metric
\begin{equation}
ds^2=L^2\fft{\mathcal F du^2+2du\,dv+(dy^1)^2+(dy^2)^2}{(y^1)^2},
\label{eq:adsbkg}
\end{equation}
along with
\begin{equation}
S=\fft3L,\qquad P=0,\qquad V_\mu=0.
\end{equation}
Setting $\mathcal F=0$ then gives AdS$_4$.  By taking, instead, $H=(y^2/L)^2$, we would
end up with an equivalent metric, but with the roles of $S$ and $P$ interchanged.
Although this background is maximally supersymmetric, only the single Killing spinor,
(\ref{eq:explicitks}), has been made explicit in this analysis. This is, therefore, a case of enhanced supersymmetry.

We leave the consideration of the condition $i_KV\ne0$ to future work as its immediate geometric meaning is not completely clear to us at this time.

\subsection{Euclidean signature}

Although supergravity is conventionally formulated in a Lorentzian signature, much of the
recent interest has been in field theories in  Euclidean signature.  At one level, this can be
thought of as introducing a Wick rotation into imaginary time.  However, the transformation is
always more subtle whenever fermions are involved.  In particular, the Majorana condition
can no longer be imposed in a four-dimensional Euclidean signature, and hence the minimal
spinor is Weyl (or a pair of symplectic Majorana-Weyl spinors if desired).

With an Euclidean signature, it is possible to take all Dirac matrices to be Hermitian.  In
four dimensions, Hermitian conjugation and transposition obey%
\footnote{Note that we could have equally well chosen charge conjugation to come with a
minus sign.}
\begin{equation}
\gamma_\mu=\gamma_\mu^\dagger,\qquad C\gamma_\mu C^{-1}=\gamma_\mu^t,
\end{equation}
where the charge conjugation matrix satisfies $C^{-1}=C^t=C^\dagger=-C$.  In this case,
Dirac conjugation is identical to Hermitian conjugation, $\bar\psi=\psi^\dagger$, while
Majorana conjugation remains $\psi^c=\psi^tC$.  Since the charge conjugation matrix $C$
has imaginary eigenvalues, we may no longer impose a Majorana condition in Euclidean
four dimensions.  Note, also, that the chirality matrix $\gamma_5$ may be defined as
$\gamma_5\equiv(1/4!)\epsilon_{\mu\nu\rho\sigma}\gamma^\mu\gamma^\nu\gamma^\rho
\gamma^\sigma$.  Unlike in the Lorentzian case, both Dirac and Majorana conjugations
preserve chirality.

We take the approach of \cite{Festuccia:2011ws} in defining the Euclidean treatment of
the old minimal supergravity theory.  In particular, note that the gravitino variation
in Lorentzian signature, (\ref{eq:deltapsi}), may be written in terms of Weyl spinors
\begin{equation}
\delta\psi_{L\mu}=[\nabla_\mu+\ft{i}6(\gamma_\mu{}^\nu-2\delta_\mu^\nu)V_\nu]\epsilon_L
-\ft16\gamma_\mu M\epsilon_R,
\label{eq:weyldeltapsi}
\end{equation}
where $M=S+iP$, and the Weyl spinors are obtained by projection,
$\epsilon_{L/R}=\fft12(1\mp\gamma_5)\epsilon$.  Since $\gamma_5$ is imaginary in the
Majorana representation, $\epsilon_L$ and $\epsilon_R$ are related by complex
conjugation, $\epsilon_R=\epsilon_L^*$.  Thus, in a Lorentzian signature, the
transformation (\ref{eq:weyldeltapsi}) corresponds to four real supersymmetries, as it
must, since we have as yet done nothing but change the notation.

The natural extension to Euclidean signature is to include both left- and right-handed projections,
so that
\begin{eqnarray}
\label{eq:eucldeltapsi}
  \delta\psi_{L\mu}&=&  \nabla_\mu{\eps_L}+\frac {1}6M\gamma_\mu{\eps_R}
    -\frac {i}3 V_\mu{\eps_L}
    -\frac {i}6 V^\nu\gamma_{\nu\mu} {\eps_L},
    \nonumber \\
 \delta\psi_{R\mu}&=&  \nabla_\mu{\eps_R}+\frac {1}6\bar M\gamma_\mu{\eps_L}
    +\frac {i}3 V_\mu{\eps_R}
    +\frac {i}6 V^\nu\gamma_{\nu\mu} {\eps_R}.
\end{eqnarray}
In doing so, we have doubled the fermionic degrees of freedom, $\psi_\mu^{\rm(Majorana)}
\to(\psi_{L\mu},\psi_{R\mu})$.  (Of course, one could introduce symplectic Majorana-Weyl
spinors, but since symplectic Majorana spinors necessarily come in pairs, this doubling of
the fermionic degrees of freedom is unavoidable in going from $3+1$ dimensions to
Euclidean four dimensions.)  In principle, for the off-shell formulation to be complete, we
would have to double the bosonic degrees of freedom as well.  Although it is not obvious
how this should be done, we may at least relax the original conditions on the bosonic
fields.  We thus take $V_\mu$ to be complex and allow $M$ and $\bar M$ to be
independent complex scalars \cite{Festuccia:2011ws}.  Ultimately, since we are not
necessarily interested in dynamical gravity, but only in supersymmetric field theories in
a curved Euclidean background, we do not need a complete formulation of Euclidean
supergravity.  Instead, we simply take (\ref{eq:eucldeltapsi}) as providing a pair of Killing
spinor equations that must be satisfied for our supersymmetric backgrounds.

In contrast to the case of new minimal supergravity \cite{Sohnius:1981tp,Sohnius:1982fw},
where the two chiral gravitino variations are independent \cite{Dumitrescu:2012ha,Klare:2012gn}, here
the $M$ and $\bar M$ terms mix the two chiralities in (\ref{eq:eucldeltapsi}).  As a result,
a complete analysis will demand the simultaneous treatment of both $\epsilon_L$ and
$\epsilon_R$.  Such an analysis was performed in \cite{Samtleben:2012gy}, where trivial
$G$-structure arises when both spinors are active, and SU(2)
structure arises in the special case when one of the chiral spinors vanish.

To see how the structure arises in the Euclidean case, we may form a complete set
of bilinears.  It is important to realize that, although we have used a Weyl notation, a single
preserved supersymmetry corresponds to a pair of Weyl spinors $(\epsilon_L,\epsilon_R$).
With this in mind, we may express the bilinears as
\begin{align}
\label{eq:inv-tensors}
&
    f^L = {\bar\eps_L}{\eps_L},
&&
    f^R = {\bar \eps_R}{\eps_R},  && \text{real positive scalars};
    \cr
&
    Q_\mu = {\bar\eps_R}\gamma_\mu{\eps_L},
&&
    K_\mu = {\eps_R^c}\gamma_\mu{\eps_L},
     && \text{complex vectors};
    \cr
&
    J^L_{\mu\nu} = i{\bar \eps_L}\gamma_{\mu\nu}{\eps_L},
&&
    J^R_{\mu\nu} = i{\bar \eps_R}\gamma_{\mu\nu}{\eps_R},
     && \text{real two forms};
    \cr
&
    \Omega^L_{\mu\nu} = {\eps^c_L}\gamma_{\mu\nu}{\eps_L},
&&
    \Omega^R_{\mu\nu} = {\eps_R^c}\gamma_{\mu\nu}{\eps_R},
     && \text{complex two forms}.
\end{align}
Note that $J^L$ and $\Omega^L$ are self-dual, while $J^R$ and $\Omega^R$ are
anti-self-dual.  Thus these bilinears contain a total of $36=(8\times 9)/2$ real components
as expected since they are built from an eight real component spinor
$\epsilon=\epsilon_L+\epsilon_R$.

The resulting structure is determined by $f^L$ and $f^R$.  Since we take $\epsilon$ to be
globally well defined and everywhere non-vanishing, we see that
$f^L+f^R=\bar\epsilon\epsilon>0$, while the chiral components are merely non-negative,
$f^L\ge0$ and $f^R\ge0$.  Locally, at least, there are three possibilities:
$i)$ both $f^L$ and $f^R$ are non-vanishing;
$ii)$ $f^R=0$; and $iii)$ $f^L=0$.  In the first case, all bilinears are non-zero, and furthermore
$K_\mu$ and $L_\mu$ span the four-dimensional space, as we indicate below.
This leads to the identification of
trivial $G$-structure \cite{Samtleben:2012gy}.  For case $ii)$, only $f^L$, $J_{\mu\nu}^L$ and
$\Omega_{\mu\nu}^L$ are present.  The normalized two-forms $J^L/f^L$ and $\Omega^L/f^L$
then define the SU(2) structure.  Case $iii)$ is similar to case $ii)$, but with the roles of left and
right interchanged.  A complete description of the structures that arise in these cases is provided
in \cite{Samtleben:2012gy}. Note, as we will exemplify below, that in many cases $f^R$ and
$f^L$ vanish at some point and the above analysis fails.

The algebraic (structure) relations may be obtained by Fierz rearrangement.  We start by
noting that $K^*_\mu K^\mu=Q^*_\mu Q^\mu=2f^Lf^R$, while all other inner products between
vectors vanish.  Furthermore, we have
\begin{equation}
K\wedge K^*\wedge Q\wedge Q^*=-4(f^Lf^R)^2*\mathbb 1.
\label{eq:orientation}
\end{equation}
So long as $f^Lf^R$ is non-vanishing, this gives rise to a trivial G-structure, as
indicated above.  The case of SU(2) structure arises when, e.g., $f^L>0$ and $f^R=0$.  To
demonstrate this, we turn to the two-forms $(J^L,\Omega^L)$ and re-express them in terms
of normalized real two-forms
\begin{equation}
\Omega^L/f^L=J^1+iJ^2,\qquad J^L/f^L=J^3.
\label{eq:Jadef}
\end{equation}
In this case, we find that
\begin{equation}
J^a_\mu{}^\nu J^b_\nu{}^\rho=-\delta^{ab}\delta_{\mu\nu}+\epsilon^{abc}J^c_{\mu\rho}.
\end{equation}
Hence the normalized two-forms $J^a$ satisfy the algebra of the
unit quaternions.  A second SU(2) structure $\tilde J^a$ arises for $(J^R,\Omega^R)$ whenever
$f^R\ne0$.

Locally, any one of the real two-forms may be used to define an almost complex structure.
When $f^Lf^R\ne0$, we see that
\begin{align}
&J^L_{\mu\nu}K^\nu=-if^LK_\mu,&&
J^R_{\mu\nu}K^\nu=-if^RK_\mu,\nonumber\\
&J^L_{\mu\nu}Q^\nu=-i f^LQ_\mu,&&
J^R_{\mu\nu}Q^\nu=if^RQ_\mu,\nonumber\\
&\Omega^L_{\mu\nu}K^\nu=0,&&
\Omega^R_{\mu\nu}K^\nu=0,\nonumber\\
&\Omega^L_{\mu\nu}K^{*\nu}=2f^LQ_\mu,&&
\Omega^R_{\mu\nu}K^{*\nu}=-2f^RQ^*_\mu,\nonumber\\
&\Omega^L_{\mu\nu}Q^\nu=0,&&
\Omega^R_{\mu\nu}Q^{*\nu}=0,\nonumber\\
&\Omega^L_{\mu\nu}Q^{*\nu}=-2f^LK_\mu,&&
\Omega^R_{\mu\nu}Q^\nu=2f^RK_\mu.
\end{align}
In fact, the two-forms can also be expressed as combinations of $K$ and $Q$
\begin{eqnarray}
f^RJ^L&=&-\ft{i}2(K\wedge K^*+Q\wedge Q^*)\nonumber\\
f^LJ^R&=&-\ft{i}2(K\wedge K^*-Q\wedge Q^*)\nonumber\\
f^R\Omega^L&=&-K\wedge Q,\nonumber\\
f^L\Omega^R&=&K\wedge Q^*.
\label{2forms}
\end{eqnarray}

Having explicitly established the algebraic relations, we now proceed to the differential relations.
Using the Killing spinor equation $\delta\psi_\mu=0$ where $\delta\psi_\mu$ is given in
(\ref{eq:eucldeltapsi}), we obtain%
\footnote{$A_{(\mu}
B_{\nu)}=\frac12\left(A_\mu B_\nu+A_\nu B_\mu\right)$ and $A_{[\mu}
B_{\nu]}=\frac12\left(A_\mu B_\nu-A_\nu B_\mu\right)$.}
\begin{eqnarray}\label{diff}
    \nabla_{\mu} f^L&=&-\ft13\Re(M^*Q_\mu) -\ft23f^L\Im V_\mu-\ft13J^L_\mu{}^\nu\Re V_\nu,
    \nn\\
    \nabla_{\mu} f^R&=& -\ft13\Re(\bar MQ_\mu)+\ft23f^R\Im V_\mu+\ft13J^R_\mu{}^\nu\Re V_\nu,
    \nn\\
    \nabla_{(\mu} K_{\nu)}&=& 0,
    \nn\\
    \nabla_{[\mu} K_{\nu]}&=& \ft16 M\Omega^R_{\mu\nu}-\ft16 \bar  M\Omega^L_{\mu\nu}
    +\ft{i}{3}\e_{\mu\nu\rho\s}V^\rho K^\s,
    \nn\\
    \nabla_{(\mu} Q_{\nu)}&=&-\ft16g_{\mu\nu}(\bar M^*f^L+Mf^R)+i(Q_{(\mu}\Re V_{\nu)}
    -\ft13g_{\mu\nu}Q^\rho\Re V_\rho),
    \nn\\
    \nabla_{[\mu} Q_{\nu]}&=&\ft{i}{6}(\bar M^*J^L_{\mu\nu}-MJ^R_{\mu\nu})
    -\ft13\epsilon_{\mu\nu\rho\sigma}(\Im V^\rho)Q^\sigma-\ft{i}3Q_{[\mu}\Re V_{\nu]},
\end{eqnarray}
along with
\begin{eqnarray}
\nabla_\mu J_{\nu\lambda}^L&=&\ft13(\epsilon_{\mu\nu\lambda\sigma}\Im(M^*Q^\sigma)
+2g_{\mu[\nu}\Im(M^*Q_{\lambda]}))
-\ft13(\epsilon_{\mu\nu\lambda\sigma}f^L\Re V^\sigma+2g_{\mu[\nu}f^L\Re V_{\lambda]})\nn\\
&&-\ft23(g_{\mu[\nu}J^L_{\lambda]\sigma}\Im V^\sigma+J^L_{\nu\lambda}\Im V_\mu
-J^L_{\mu[\nu}\Im V_{\lambda]}),\nn\\
\nabla_\mu J_{\nu\lambda}^R&=&\ft13(\epsilon_{\mu\nu\lambda\sigma}\Im(\bar MQ^\sigma)
-2g_{\mu[\nu}\Im(\bar MQ_{\lambda]}))
-\ft13(\epsilon_{\mu\nu\lambda\sigma}f^R\Re V^\sigma-2g_{\mu[\nu}f^R\Re V_{\lambda]})\nn\\
&&+\ft23(g_{\mu[\nu}J^R_{\lambda]\sigma}\Im V^\sigma+J^R_{\nu\lambda}\Im V_\mu
-J^R_{\mu[\nu}\Im V_{\lambda]}),\nn\\
\nabla_\mu\Omega_{\nu\lambda}^L&=&-\ft13(\epsilon_{\mu\nu\lambda\sigma}MK^\sigma
+2g_{\mu[\nu}MK_{\lambda]})+\ft{2i}3(g_{\mu[\nu}\Omega^L_{\lambda\sigma}V^\sigma
+\Omega^L_{\nu\lambda}V_\mu-\Omega^L_{\mu[\nu}V_{\lambda]}),\nn\\
\nabla_\mu\Omega_{\nu\lambda}^R&=&-\ft13(\epsilon_{\mu\nu\lambda\sigma}\bar MK^\sigma
-2g_{\mu[\nu}\bar MK_{\lambda]})-\ft{2i}3(g_{\mu[\nu}\Omega^R_{\lambda\sigma}V^\sigma
+\Omega^R_{\nu\lambda}V_\mu-\Omega^R_{\mu[\nu}V_{\lambda]}).
\label{eq:diff2}
\end{eqnarray}

\subsubsection{SU(2) structure}

We now briefly consider the case of SU(2) structure.  Assuming, say, that $f^R=0$, the differential
identities (\ref{diff}) and (\ref{eq:diff2}) reduce to
\begin{eqnarray}
\nabla_{\mu} f^L&=&-\ft23f^L\Im V_\mu-\ft13J^L_\mu{}^\nu\Re V_\nu,\nn\\
\nabla_\mu J_{\nu\lambda}^L&=&
-\ft13(\epsilon_{\mu\nu\lambda\sigma}f^L\Re V^\sigma+2g_{\mu[\nu}f^L\Re V_{\lambda]})\nn\\
&&-\ft23(g_{\mu[\nu}J^L_{\lambda]\sigma}\Im V^\sigma+J^L_{\nu\lambda}\Im V_\mu
-J^L_{\mu[\nu}\Im V_{\lambda]}),\nn\\
\nabla_\mu\Omega_{\nu\lambda}^L&=&
\ft{2i}3(g_{\mu[\nu}\Omega^L_{\lambda]\sigma}V^\sigma
+\Omega^L_{\nu\lambda}V_\mu-\Omega^L_{\mu[\nu}V_{\lambda]}),
\end{eqnarray}
along with the constraint $\bar M=0$.  Since in this case $\epsilon_R=0$, the Killing spinor
equation only involves a single chiral spinor, and the analysis is similar to that of Sec.~3 of
\cite{Dumitrescu:2012ha}.

\subsubsection{Trivial structure}

For $f^Lf^R\ne0$, the differential identities (\ref{diff}) imply the existence of two Killing vectors,
$K$ and $K^*$.  We compute
\begin{equation}
[K,K^*]=2i\Im(K^\mu\nabla_\mu K^*)=-2iL,
\label{eq:kvalg}
\end{equation}
where
\begin{equation}
L=\Im(\lambda Q),\qquad\lambda=\ft13(2Q_\lambda^*\Im V^\lambda+f^L\bar M-f^RM^*).
\label{eq:Ldef}
\end{equation}
As a result, there are two possibilities, depending on whether $\lambda$ vanishes or not.  For
$\lambda=0$, the two Killing vectors commute, and
we end up with a torus fibration.  When this quantity is non-vanishing, on the other hand,
we obtain a third Killing vector $L$, such that $K$, $K^*$ and $L$ satisfy an SU(2) algebra.
(Note that $\Re K$, $\Im K$ and $L$ are mutually orthogonal.)

Before proceeding with an analysis of these two cases, we first note that a direct computation
of the Nijenhuis tensor demonstrates that $J^3=J^L/f^L$ defined in (\ref{eq:Jadef}) is an
integrable complex structure, so long as $f^L$ is non-vanishing.  Similarly, $\tilde J^3=J^R/f^R$
is integrable and well-defined whenever $f^R$ is non-vanishing.  This proves that, at least locally,
we can introduce complex coordinates on the four-dimensional Euclidean manifold using either
one of the two complex structures $J^3$ and $\tilde J^3$.  It is important to note, however, that
even though we are considering the case $f^Lf^R\ne0$, we must allow for the possibility that
$f^L$ (or $f^R$) may vanish at isolated points or along some curves.  In this case, $J^3$
(and similarly $\tilde J^3$) would not be globally defined, and hence the Euclidean manifold
is not necessarily complex.  An example of this is the case of $\mathrm{S}^4$, which admits a maximal
set of Killing spinors, but which is not a complex manifold.

Returning to the algebra of Killing vectors, (\ref{eq:kvalg}), we first consider the case when
$K$ and $K^*$ commute.  By introducing real coordinates $y_1$ and $y_2$ with
\begin{equation}
K = \frac{\pd}{\pd y_1} + i \frac{\pd}{\pd y_2},
\end{equation}
we can write the metric as a torus fibration over a two-dimensional manifold $X$
\begin{equation}
    ds^2 =
    ds^2_X + e^{2A}\left(dy_1+w\right)^2
    +e^{2B}\left(dy_2+\alpha\, dy_1+\wt w\right)^2,
\end{equation}
where $w$, $\wt w$ are one forms and $A$ and $B$ are scalars on $X$.  The normalization of
$K$ then demands that $e^{2A}=e^{2B}=f^Lf^R$ as well as $\alpha=0$.  We thus
write
\begin{equation}
    ds^2 =
    f^2\left[e^{2\sigma}(dx_1^2+dx_2^2)+ \left(dy_1+w\right)^2
    +\left(dy_2+\wt w\right)^2\right],
\label{eq:T2isomet}
\end{equation}
where $f^2=f^Lf^R$, and we have made explicit the metric on $X$.  This metric admits
a natural vierbein basis
\begin{equation}
e^1=f(dy_1+w),\qquad e^2=f(dy_2+\tilde w),\qquad e^3=fe^\sigma dx_1,\qquad
e^4=fe^\sigma dx_2.
\end{equation}
In this case, the vectors $K$ and $Q$ take the form
\begin{equation}
K=f(e^1+ie^2),\qquad Q=f(e^3+ie^4).
\label{eq:KQtorus}
\end{equation}

Our goal is now to apply the differential identities (\ref{diff}) to solve for the auxiliary fields
$M$, $\bar M$ and $V$.  Since we are working with trivial structure, the two-forms are
completely determined by $K$ and $Q$, as in (\ref{2forms}).  As a result, the two-form differential
identities (\ref{eq:diff2}) will not introduce any further conditions, and hence will not need to be
examined.  To proceed, we decompose $V$ along $K$ and $Q$
\begin{equation}
V=\alpha Q+\beta Q^*+\gamma K+\delta K^*,
\label{eq:acomp}
\end{equation}
where for now the quantities $(\alpha,\beta,\gamma,\delta)$ are complex.  As in the Lorentzian
case, we restrict to the case $i_KV=i_KV^*=0$.  This immediately sets $\gamma=\delta=0$.
Furthermore, this restriction combined with the first two equations in (\ref{diff}) demonstrates
that $i_Kdf^L=i_Kdf^R=0$, so that the isometry of the metric extends to $f^L$ and $f^R$.

Using the decomposition (\ref{eq:acomp}) along with the form of $K$ and $Q$ given in
(\ref{eq:KQtorus}), we find that the $df^L$ and $df^R$ identities give rise to the conditions
\begin{equation}
M=-\fft3{f^2}e^{-\sigma}\partial f^L-if^L(3\alpha^*-\beta),\qquad
\bar M=-\fft3{f^2}e^{-\sigma}\bar\partial f^R+if^R(-\alpha+3\beta^*),
\end{equation}
where we have defined
\begin{equation}
\partial\equiv\fft\partial{\partial x_1}+i\fft\partial{\partial x_2},\qquad
\bar\partial\equiv\fft\partial{\partial x_1}-i\fft\partial{\partial x_2}.
\end{equation}
We now turn to the $dK$ identity in (\ref{diff}).  The components of this identity on $T^2$ give
\begin{equation}
M=-\fft{3f^L}{f^3}e^{-\sigma}\partial f-2if^L\beta,\qquad
\bar M=-\fft{3f^R}{f^3}e^{-\sigma}\bar\partial f+2if^R\alpha,
\end{equation}
while the components on the base $X$ give
\begin{equation}
dw=d\tilde w=0.
\end{equation}
This demonstrates that $w$ and $\tilde w$ are trivial, and can be removed by a gauge
transformation.  Finally, the $dQ$ identity gives
\begin{equation}
M=-\fft{3f^L}{2f^4}e^{-2\sigma}\partial(f^2e^\sigma)+\fft{i}2f^L(3\alpha^*-\beta),\qquad
\bar M=-\fft{3f^R}{2f^4}e^{-2\sigma}\bar\partial(f^2e^\sigma)+\fft{i}2f^R(\alpha-3\beta^*).
\end{equation}

Taking $f^L$ and $f^R$ as two independent functions specifying the background, we
may solve the above equations for the auxiliary fields
\begin{eqnarray}
M&=&-\fft{f^L}{f^2}e^{-\sigma}\left(\fft{\partial f^R}{f^R}+2\fft{\partial f^L}{f^L}+\partial\sigma\right),\nn\\
\bar M&=&-\fft{f^R}{f^2}e^{-\sigma}\left(2\fft{\bar\partial f^R}{f^R}+\fft{\bar\partial f^L}{f^L}
+\bar\partial\sigma\right),
\end{eqnarray}
and
\begin{equation}
V=\alpha Q+\beta Q^*=f[\alpha(e^3+ie^4)+\beta(e^3-ie^4)],
\end{equation}
where
\begin{eqnarray}
\alpha=\fft{i}{4f^2}e^{-\sigma}\left(\fft{\bar\partial f^R}{f^R}-\fft{\bar\partial f^L}{f^L}
+2\bar\partial\sigma\right),\nn\\
\beta=\fft{i}{4f^2}e^{-\sigma}\left(\fft{\partial f^R}{f^R}-\fft{\partial f^L}{f^L}
-2\partial\sigma\right).
\end{eqnarray}
To summarize, supersymmetric backgrounds admitting two commuting Killing vectors
may be written in the form (\ref{eq:T2isomet}), with $f=f^Lf^R$, and with auxiliary fields
given above.  Such backgrounds are parameterized by two arbitrary functions,
$f^L(x_1,x_2)$ and $f^R(x_1,x_2)$.  Note that we can verify that this solution yields
$\lambda=0$, so that $L=0$, where $L$ is defined in (\ref{eq:Ldef}).  Thus $K$ and $K^*$
commute, as they must by construction in this case.

We now consider the case when $K$, $K^*$ and $L$ satisfy an SU(2) algebra.  Following
\cite{Dumitrescu:2012ha}, but taking the case of old minimal supergravity, we note that the
isometry may be made explicit by taking a metric of the form
\begin{equation}
ds^2=d\xi^2+f(\xi)^2(\sigma_1^2+\sigma_2^2+\sigma_3^2),
\label{eq:su2met}
\end{equation}
where $\sigma^a$ are a set of left-invariant one-forms satisfying
$d\sigma^a=-\fft12\epsilon_{abc}\sigma^b\wedge\sigma^c$.  Introducing a set of vierbeins
\begin{equation}
e^a=f\sigma^a,\qquad e^4=d\xi,
\label{eq:su2vier}
\end{equation}
and making note of (\ref{eq:orientation}), we may specialize the vectors $K$ and $Q$ to this
metric
\begin{equation}
K=f(e^1+ie^2),\qquad Q=f(e^3+ie^4).
\label{eq:frame}
\end{equation}
Note that we have chosen a normalization $f^Lf^R=f^2$.

As in the torus fibration case, the solution is determined by examining the differential
identities (\ref{diff}).  Using the same decomposition (\ref{eq:acomp}), and again restricting
to $\gamma=\delta=0$, but this time choosing the frame (\ref{eq:frame}) with vierbeins
given in (\ref{eq:su2vier}), we find that the $df^L$ and $df^R$ identities give
\begin{equation}
M=-3i\fft{(f^L)'}f-if^L(3\alpha^*-\beta),\qquad
\bar M=3i\fft{(f^R)'}f+if^R(-\alpha+3\beta^*).
\end{equation}
Similarly, the $dK$ identity gives
\begin{equation}
M=3i\fft{f^L}{f^2}(\ft12-f')-2if^L\beta,\qquad
\bar M=3i\fft{f^R}{f^2}(\ft12+f')+2if^R\alpha,
\end{equation}
and the $dQ$ identity gives
\begin{equation}
M=3i\fft{f^L}{f^2}(\ft12-f')+\fft{i}2f^L(3\alpha^*-\beta),\qquad
\bar M=3i\fft{f^R}{f^2}(\ft12+f')+\fft{i}2f^R(\alpha-3\beta^*).
\end{equation}
These equations can be solved for
\begin{eqnarray}
M&=&-i\fft{f^L}f\left(\fft{(f^R)'}{f^R}+2\fft{(f^L)'}{f^L}-\fft1f\right),\nn\\
\bar M&=&i\fft{f^R}f\left(2\fft{(f^R)'}{f^R}+\fft{(f^L)'}{f^L}+\fft1f\right),\nn\\
-\beta=\alpha&=&\fft1{4f}\left(\fft{(f^R)'}{f^R}-\fft{(f^L)'}{f^L}-\fft1f\right).
\label{eq:MbMa}
\end{eqnarray}
Note, in particular, that with this gauge choice $M$ and $\bar M$ are pure imaginary.
Furthermore, according to (\ref{eq:acomp}), the vector $V$ is given by
\begin{equation}
V=2i\alpha\,\Im Q=\fft{i}2\left(\fft{(f^R)'}{f^R}-\fft{(f^L)'}{f^L}-\fft1f\right) e^4.
\label{eq:avec}
\end{equation}

As a result, we have seen that solutions with $S^3$ isometry may be specified by two
functions $f^L(\xi)$ and $f^R(\xi)$.  The metric is given by (\ref{eq:su2met}) with
$f=f^Lf^R$, and the auxiliary fields are given in (\ref{eq:MbMa}) and (\ref{eq:avec}).
For self-consistency, we can check that this solution yields $\lambda=i$, so that the Killing
vector $L$ defined in (\ref{eq:Ldef}) is simply $L=\Im(iQ)=\Re Q=fe^3$.  The three Killing
vectors of this solution are thus $K=f^2(\sigma^1+i\sigma^2)$ and $L=f^2\sigma^3$.

\subsubsection{Some global aspects of Euclidean supersymmetry}

At this point, we wish to reemphasize the fact that the above construction may fail to yield an
almost complex structure. The existence of supersymmetry requires a nowhere vanishing Killing spinor. However, in our construction we have split the spinor into left- and right-handed chiral
components. The norm that is nowhere vanishing is the norm of $\epsilon$:
\be
\bar{\epsilon} \epsilon= f^L+f^R>0.
\ee
Thus, it is only the sum of $f^L$ and $f^R$ in (\ref{eq:inv-tensors}) that is nowhere vanishing.
A simple way of stating the above equation is by saying that the nonvanishing norm ``bounces''
among the left and right parts of the spinor. Under this condition, the natural candidate for an
almost complex structure
\be
J_{\mu\nu}=\fft{J^L_{\mu\nu}}{f^L}=
i\frac{\bar{\epsilon}_L \gamma_{\mu\nu}\epsilon_L}{\bar{\epsilon}_L \epsilon_L},
\ee
is not everywhere well-defined. Similarly normalized, $\tilde J=J^{R}_{\mu\nu}/f^R$  cannot
be everywhere well-defined.

In the next section where we present explicit solutions we will illustrate that the round $\mathrm{S}^4$
fails to admit an almost complex structure in precisely this ``bouncing spinor'' way.  Here we
simply note that for the $S^3$ isometry case, we have
\begin{equation}
J=\fft1{2if^Lf^R}(K\wedge K^*+Q\wedge Q^*)=-(e^1\wedge e^2+e^3\wedge e^4)
=-(f^2\sigma^1\wedge\sigma^2+f\sigma^3\wedge d\xi).
\end{equation}
This degenerates whenever $f=0$.

\subsection{Explicit Solutions}

Here we focus on supersymmetric backgrounds with $S^3$ isometry, with metric given by
(\ref{eq:su2met}) and auxiliary fields given by (\ref{eq:MbMa}) and (\ref{eq:avec}).  These
backgrounds include the round and distorted $\mathrm{S}^4$, as well as $\mathbb R\times S^3$.

\subsubsection{Halving the supersymmetry}

Before examining some solutions, it is worth noting the explicit form of the Killing spinor
$\epsilon=(\epsilon_L,\epsilon_R)$.  Using the standard parametrization for the left-invariant
one-forms
\begin{eqnarray}
\sigma^1&=&\sin\psi d\theta-\cos\psi\sin\theta d\phi,\nn\\
\sigma^2&=&\cos\psi d\theta+\sin\psi\sin\theta d\phi,\nn\\
\sigma^3&=&d\psi+\cos\theta d\phi,
\end{eqnarray}
along with the vierbein basis $e^a=f\sigma^a$, $e^4=d\xi$, we obtain
\begin{eqnarray}
\nabla_1&=&\fft1f\left(\sin\psi\fft\partial{\partial\theta}-\cos\psi\csc\theta\fft\partial{\partial\phi}
+\cos\psi\cot\theta\fft\partial{\partial\psi}\right)-\fft1{4f}\gamma^{23}+\fft{f'}{2f}\gamma^{14},\nn\\
\nabla_2&=&\fft1f\left(\cos\psi\fft\partial{\partial\theta}+\sin\psi\csc\theta\fft\partial{\partial\phi}
-\sin\psi\cot\theta\fft\partial{\partial\psi}\right)+\fft1{4f}\gamma^{13}+\fft{f'}{2f}\gamma^{24},\nn\\
\nabla_3&=&\fft1f\fft\partial{\partial\psi}-\fft1{4f}\gamma^{12}+\fft{f'}{2f}\gamma^{34},\nn\\
\nabla_4&=&\fft\partial{\partial\xi},
\end{eqnarray}
when acting on a spinor $\epsilon$.

If we take $\epsilon$ to be independent of the $S^3$ coordinates, the Killing spinor equations
following from (\ref{eq:eucldeltapsi}) reduce to
\begin{eqnarray}
0=D_1\epsilon_L&=&\left(-\fft1{2f}(\ft12-f')+\fft{i}6V^4\right)\gamma^{23}\epsilon_L
+\fft16M\gamma_1\epsilon_R,\nn\\
0=D_1\epsilon_R&=&\left(-\fft1{2f}(\ft12+f')+\fft{i}6V^4\right)\gamma^{23}\epsilon_R
+\fft16\bar M\gamma_1\epsilon_L,\nn\\
0=D_4\epsilon_L&=&\left(\fft\partial{\partial\xi}-\fft{i}3V^4\right)\epsilon_L
+\fft16M\gamma_4\epsilon_R,\nn\\
0=D_4\epsilon_R&=&\left(\fft\partial{\partial\xi}+\fft{i}3V^4\right)\epsilon_R
+\fft16\bar M\gamma_4\epsilon_L.
\end{eqnarray}
Note that we have taken the vector $V$ to point only along the $e^4$ direction, as implied by
(\ref{eq:avec}).  Substituting in for $f=\sqrt{f^Lf^R}$, $M$, $\bar M$ and $V^4$, we then
arrive at the solution to the Killing spinor equations
\begin{equation}
\epsilon=(\epsilon_L,\epsilon_R)\qquad\mbox{where}\qquad
\epsilon_L=\sqrt{f^L}\epsilon_L^0,\qquad
\epsilon_R=-i\sqrt{f^R}\gamma_4\epsilon_L^0.
\label{eq:S3ks}
\end{equation}
Since this is based on a constant chiral spinor $\epsilon_L^0$, we see that generically only half
of the supersymmetries are preserved.

Note that in the torus fibration case, with metric given by (\ref{eq:T2isomet}), the Killing spinor
has an identical form as (\ref{eq:S3ks}), but must also satisfy the projection
\begin{equation}
\gamma^{34}\epsilon_L^0=i\epsilon_L^0.
\end{equation}
Generically, this preserves a quarter of the supersymmetries.  However, this gets enhanced to
a half of the supersymmetries in the case where $f^L$ and $f^R$ are independent of one of
the base coordinates (say $x_1$).

\subsubsection{The hyperbolic sphere $\mathbb{H}^4$}\label{Hyper4sphere}
Let us consider an explicit solution of the torus fibered Ansatz  (\ref{eq:T2isomet}). We take $\sigma=0$, and $f^L=f^R$. This choice leads to $\alpha=\beta=0$ which means vanishing gauge field. If we further take $f^L=\ell/x_2$ we have
\bea
ds^2&=&\frac{\ell^2}{x_2^2}\bigg[dx_1^2 +dx_2^2+dy_1^2+dy_2^2\bigg], \nonumber \\
M&=&\frac{3i}{\ell}, \qquad \bar{M}=-\frac{3i}{\ell}.
\eea

\subsubsection{The round $\mathrm{S}^4$}\label{Round4sphere}

The round four-sphere may be obtained by taking
\begin{equation}
f^L=\cos^2(\xi/2),\qquad f^R=\sin^2(\xi/2),
\label{eq:RoundS4}
\end{equation}
so that $f^2=f^Lf^R=\fft14\sin^2\xi$.  The metric (\ref{eq:su2met}) is then
\begin{equation}
ds^2=d\xi^2+\ft14\sin^2\xi(\sigma_1^2+\sigma_2^2+\sigma_3^2),
\end{equation}
and may be identified as that of the round $\mathrm{S}^4$.  Inserting (\ref{eq:RoundS4}) into
(\ref{eq:MbMa}), we find the expected result for the auxiliary fields
\begin{equation}
M=\bar M=3i,\qquad V=0.
\end{equation}

Although the round $\mathrm{S}^4$ is maximally symmetric, there is a preferred Killing spinor used in
the invariant tensor analysis, namely the one given in (\ref{eq:S3ks}).  For (\ref{eq:RoundS4}), this
spinor has the form
\begin{equation}
\epsilon=(\cos(\xi/2)-i\gamma^4\sin(\xi/2))\epsilon_L^0=e^{-\fft{i}2\xi\gamma^4}\epsilon_L^0.
\end{equation}
At the expense of being pedantic, but with the hopes of completely clarifying the subtle
topological point above, we note that the norms of $\epsilon_R$ and $\epsilon_L$ vanish
at $\xi=0$ (the north pole) and $\xi=\pi$ (the south pole), respectively
\begin{equation}
\epsilon_R^\dagger\epsilon_R=f^R=\sin^2(\xi/2),\qquad
\epsilon_L^\dagger\epsilon_L=f^L=\cos^2(\xi/2).
\end{equation}
However, the Dirac norm is in fact constant
\begin{equation}
\epsilon^\dagger\epsilon=f^R+f^L=1.
\end{equation}
Thus, while the Killing spinor is in fact globally defined and everywhere non-vanishing, its Weyl
components will vanish at the north and south poles.  The would-be complex structure
$J=J^L/f^L$ is thus well defined in the northern hemisphere, but breaks down at the south pole,
while the would-be complex structure $\tilde J=J^R/f^R$ is well defined in the southern
hemisphere, but breaks down at the north pole.  This is the supergravity counterpart to the
well-known fact that there is no almost complex structure on the round $\mathrm{S}^4$.

\subsubsection{The $\mathbb \mathbb{R}\times \mathrm{S}^3$ case}

By taking constant
\begin{equation}
f^L=\fft\ell2,\qquad f^R=\fft\ell2,
\end{equation}
we end up with a direct product metric
\begin{equation}
ds^2=d\xi^2+\ft14\ell^2(\sigma_1^2+\sigma_2^2+\sigma_3^2).
\end{equation}
In this case, the auxiliary fields take the values
\begin{equation}
M=\bar M=\fft{2i}\ell,\qquad V=-\fft{i}\ell d\xi.
\end{equation}
%

\subsubsection{The squashed $\mathrm{S}^4$ as a supersymmetric background}

Having looked at the round $\mathrm{S}^4$, it is natural to consider squashed backgrounds that
preserve the topology of $\mathrm{S}^4$.  Since the solution (\ref{eq:MbMa}) is specified by two
arbitrary functions $f^L$ and $f^R$, it may at first appear that no additional restrictions
are needed.  However, while this is true from a local analysis, we must additionally impose
regularity at the north and south poles of the $\mathrm{S}^4$.  Rewriting the metric (\ref{eq:su2met}) as
\begin{equation}
ds^2=d\xi^2+(f^Lf^R)(\sigma_1^2+\sigma_2^2+\sigma_3^2),
\end{equation}
we see that regularity at the north pole (i.e.\ as $\xi\to0$) requires
\begin{equation}
f^Lf^R=\ft14\xi^2+\cdots,\qquad\mbox{as }\xi\to0.
\end{equation}
However, since $f^L$ and $f^R$ cannot simultaneously vanish (since otherwise the
Killing spinor $\epsilon$ would vanish), we are led to the expansion
\begin{equation}
f^L=\lambda^2+\mathcal O(\xi^2),\qquad
f^R=\ft14(\xi/\lambda)^2(1+\mathcal O(\xi^2)),\qquad\mbox{as }\xi\to 0,
\label{eq:fLfRcond}
\end{equation}
where $\lambda$ is a constant.  The absence of the linear terms in (\ref{eq:fLfRcond}) is
not imposed by the geometry, but rather arises when we demand that the auxiliary fields
$M$, $\bar M$ and $V$ are well-behaved at the north pole.

It is clear that the round $\mathrm{S}^4$, given by (\ref{eq:RoundS4}), satisfy these regularity
conditions, as
\begin{equation}
f^L=\cos^2(\xi/2)=1-\ft14\xi^2+\cdots,\qquad
f^R=\sin^2(\xi/2)=\ft14\xi^2(1-\ft1{12}\xi^2+\cdots).
\end{equation}
Of course, in order for the sphere to close at the south pole, we also need $f$ to vanish
as $\xi\to\xi_0$ (where we take $\xi_0$ to denote the location of the south pole).  The
regularity condition at $\xi_0$ is similar to (\ref{eq:fLfRcond}), but with the roles of $f^L$
and $f^R$ interchanged
\begin{equation}
f^L=\ft14((\xi-\xi_0)/\bar\lambda)^2(1+\mathcal O((\xi-\xi_0)^2)),\qquad
f^R=\bar\lambda^2+\mathcal O((\xi-\xi_0)^2),\qquad\mbox{as }\xi\to\xi_0.
\end{equation}
(One may wonder whether it is possible for, say, $f^L$ to vanish at both poles, while
$f^R$ remains non-zero.  However, it is possible to show that this cannot happen, as the
northern and southern hemisphere solutions cannot be patched together in this case.)
As an example, we may write down a simple polynomial solution that is regular at both
the north and south poles
\begin{equation}
f^L=\fft{\xi_0}{2\sqrt2}(2-(1-\xi/\xi_0)^2)(1-\xi/\xi_0)^2,\qquad
f^R=\fft{\xi_0}{2\sqrt2}(\xi/\xi_0)^2(2-(\xi/\xi_0)^2),
\end{equation}
where $\xi\in[0,\xi_0]$.
This solution has non-constant $M$ and $\bar M$, as well as a vector field $V$ turned on.
An alternate approach to squashing the $\mathrm{S}^4$ is detailed in Appendix~\ref{app:S4}.

\subsubsection{The half supersymmetric $\mathbb{R}^4$}
As a degenerate case of the previous section, we may consider what happens when $f$ vanishes at a single
point, so that $\xi$ can take values on a half-open interval.  The result is then a `cigar'
geometry in general.  However, if we take
\begin{equation}
f^L=1,\qquad f^R=\ft14\xi^2,
\end{equation}
we end up with a metric on flat $\mathbb R^4$.  This solution is somewhat curious,
since the auxiliary fields computed from (\ref{eq:MbMa}) are
\begin{equation}
M=0,\qquad\bar M=3i,\qquad V=0.
\end{equation}
Since $\bar M\ne0$, this background preserves only half of the supersymmetries,
with a Killing spinor given by $\epsilon_L=0$ and $\nabla_\mu\epsilon_R=0$.

\section{All supersymmetric backgrounds of new minimal supergravity}\label{Sec:NewMin}

It is instructive to compare the above supersymmetry analysis with the corresponding case of
new minimal supergravity \cite{Sohnius:1981tp,Sohnius:1982fw}.  Since the Euclidean
analysis was performed in \cite{Dumitrescu:2012ha,Klare:2012gn}, we limit our discussion to Lorentzian
signature.  The fields of new minimal supergravity consist of the graviton $g_{\mu\nu}$
and gravitino $\psi_\mu$ along with two auxiliary fields $A_\mu$ and $V_\mu$ (with
$\nabla^\mu V_\mu=0$).

The gravitino variation is given by
\begin{equation}
\delta\psi_\mu=\mathcal D_\mu\epsilon\equiv[\nabla_\mu+i\gamma_5A_\mu
-\ft{i}2(\gamma_\mu{}^\nu-2\delta_\mu^\nu)\gamma_5V_\nu]\epsilon,
\end{equation}
where the spinors are taken to be Majorana.  Formally this has a similar structure to
the corresponding variation in old minimal supergravity, (\ref{eq:deltapsi}), except that
the complex scalar $S+iP$ is no longer present.

Just as in the above analysis, the presence of a Killing spinor implies the structure given by
(\ref{eq:algident}) and (\ref{eq:Xdef}).  In this case, however, the differential identities take the
form
\begin{eqnarray}
\nabla_\mu K_\nu&=&-\epsilon_{\mu\nu}{}^{\rho\sigma}V_\rho K_\sigma,\nonumber\\
\nabla_\mu J_{\nu\lambda}&=&2(-(A_\mu+V_\mu)*J_{\nu\lambda}
+*J_{\mu[\nu}V_{\lambda]}-g_{\mu[\nu}*J_{\lambda]\alpha}V^\alpha),
\label{eq:diffident2}
\end{eqnarray}
so that, in particular
\begin{equation}
\nabla_{(\mu}K_{\nu)}=0,\qquad dK=2i_K*V,\qquad dJ=-2*J\wedge A,\qquad d*J=2J\wedge A.
\end{equation}
We may construct a supersymmetric background by taking an identical metric ansatz as
(\ref{eq:metans}).  In this case, we find
\begin{eqnarray}
\label{eq:newminback}
V_m&=&\ft12H^{-1/2}\hat\epsilon_m{}^n\partial_nH,\nonumber\\
A_m&=&\ft12H^2[-X_m\hat\epsilon^{np}\partial_n(H^{-3/2}X_p)
+\hat\epsilon_{mn}X^n\hat\nabla^p(H^{-3/2}X_p)],\nonumber\\
A_++\ft32V_+&=&-\ft12H\hat\epsilon_{mn}X^m\partial_uX^n,
\end{eqnarray}
where we have assumed from the start that $A_-=V_-=0$.  The Killing spinor satisfies the
projection $\gamma^+\epsilon=0$, and is given in conformal gauge by (\ref{eq:explicitks}),
{\it except} that the projection $\gamma^1\epsilon_0=\epsilon_0$ is no longer required.
In this case, generic supersymmetric backgrounds preserve two of the four supersymmetries.

\subsection{AdS$_4$ in new minimal supergravity}
It is interesting to see that new minimal supergravity can also lead to an AdS$_4$ background.
Taking the same configuration as (\ref{eq:adsbkg}), we find
\begin{eqnarray}
&&ds^2=L^2\fft{\mathcal F du^2+2du\,dv+(dy^1)^2+(dy^2)^2}{(y^1)^2},\nonumber\\
&&V=-\fft1{y^1}dy^2,\qquad A=\fft3{2y^1}dy^2.
\end{eqnarray}
While the metric is maximally symmetric, the auxiliary fields clearly break this symmetry.  As
a result, the background only preserves two of the four supersymmetries.  This may be seen
by examining the integrability condition arising from
\begin{eqnarray}
[\mathcal D_\mu,\mathcal D_\nu]&=&\ft14[R_{\mu\nu}{}^{\rho\sigma}
+2(\delta_\mu^\rho\delta_\nu^\sigma V^2
-2\delta_{[\mu}^\rho V_{\nu]}V^\sigma)
+2\epsilon_{\alpha\beta}{}^{\rho\sigma}\delta_{[\mu}^\alpha\nabla_{\nu]}V^\beta]
\gamma_{\rho\sigma}\nonumber\\
&&+2i\gamma_5(\partial_{[\mu}A_{\nu]}+\partial_{[\mu}V_{\nu]}).
\label{eq:newintr}
\end{eqnarray}
For AdS$_4$ with radius $L$, the first two terms in (\ref{eq:newintr}) cancel completely
since $V^2=1/L^2$ and $R_{\mu\nu}{}^{\rho\sigma}=-(1/L^2)(\delta_\mu^\rho\delta_\nu^\sigma
-\delta_\nu^\rho\delta_\mu^\sigma)$.  The remaining terms, schematically $V^2+\nabla V
+d(A+V)$, can then be grouped together to multiply an overall $\gamma^+$.  Hence we conclude
that the projection $\gamma^+\epsilon=0$ cannot be removed for this background.

The auxiliary fields $A$ and $V$ yield particular couplings to the supersymmetric field theory of interests. We postpone the discussion of the specific form of the field theory Lagrangian for future work.

It is worth emphasizing that we have not used the equations of motion, and therefore the class of metrics allowing for a supersymmetric background is rather wide. In fact, the main requirement is the existence of a null Killing vector. This is in stark contrast to the situation when a classification of the solutions is sought \cite{Tod:1983pm,Gauntlett:2002nw}. Note that recently some BPS Lifshitz and Schr\"odinger solutions in $D=4, {\cal N}=1$ off-shell supergravity have been constructed \cite{Lu:2012am,Liu:2012cv,Lu:2012cz}

\section{Conclusions}\label{Sec:Conclusions}

In this paper we have discussed the conditions for rigid supersymmetry arising in ${\cal N}=1$ off-shell supergravity. We have focused on the old minimal supergravity, given recent work in the new minimal supergravity \cite{Dumitrescu:2012ha,Klare:2012gn}. One of the main results is a complete and explicit description of {\it all  supersymmetric backgrounds of both the old minimal and the new minimal supergravity in Lorentzian signature admitting a hypersurface orthogonal Killing vector}. As follows from equations (\ref{eq:metans}), (\ref{eq:cgaux}) and (\ref{eq:newminback}), all fields and the metric are fully determined in terms of three functions. It is interesting to highlight that given the auxiliary fields of the new minimal supergravity one would not naively expect a maximally symmetric space to solve the corresponding gravitino variations, as intuition dictates that having a non-vanishing vector leads to a preferred direction in space therefore breaking the symmetry. We have verified that the intuition is misleading, as the two vectors of new minimal supergravity can conspire in a precise way so as to lead, in the Lorentzian case, to a solution with $AdS_4$ albeit with less preserved supersymmetry that in the old minimal model.

There are a few interesting problems that we would like to highlight. First, there is the natural question of localization for the supersymmetric field theories on compact rigid supersymmetric backgrounds discussed here. It will be particularly interesting to understand localization in the case of the squashed four-sphere explicitly constructed here.  Another natural question is the study of rigid supersymmetry in the case of theories with eight supercharges. In particular, recent work related to the five-dimensional theories \cite{Hosomichi:2012ek,Kawano:2012up,Kallen:2012va,Kim:2012av} should be revisited under the framework of rigid supersymmetry.  We hope to return to some of this issues in the future.

\section*{Acknowledgments}
We are very grateful to Kentaro Hanaki for various discussions and clarifications about off-shell supersymmetry.
This research was supported in part by the Department
of Energy under grant DE-FG02-95ER40899 to the University of Michigan.

\appendix

\section{Toward arbitrary dimensions: Comments on generic gravitino variations}\label{App:Integrability}

It is well known that maximally symmetric spaces admit a complete set of Killing spinors
$\epsilon$ satisfying the Killing spinor equation $D_\mu\epsilon=0$ where
\begin{equation}
D_\mu\equiv\nabla_\mu+m\gamma_\mu.
\label{eq:Dorig}
\end{equation}
(For simplicity we focus on Dirac spinors in $d$ dimensions, although
the results apply more generally as well.)  In particular, taking $m$ to be constant yields
the integrability constraint
\begin{equation}
[D_\mu,D_\nu]=\ft14[R_{\mu\nu}{}^{\rho\sigma}
+4m^2(\delta_\mu^\rho\delta_\nu^\sigma-\delta_\mu^\sigma\delta_\nu^\rho)]\gamma_{\rho\sigma},
\label{eq:origintr}
\end{equation}
which is solved for maximally symmetric backgrounds satisfying
\begin{equation}
R_{\mu\nu\rho\sigma}=-4m^2(g_{\mu\rho}g_{\nu\sigma}-g_{\mu\sigma}g_{\nu\rho}).
\label{eq:Rmax}
\end{equation}
(Note that we allow $m^2$ to have either sign.)
It is furthermore clear that such backgrounds preserve a maximum set of supersymmetries  as no projection condition on the spinor is required.

Our goal is to deform a maximally symmetric background in such a way that it will continue to
preserve at least a fraction of the original supersymmetries.  As long as we do not modify the
Killing spinor equation built out of (\ref{eq:Dorig}), the set of possible deformations is rather
restrictive.  To see this, we may multiply (\ref{eq:origintr}) on the left by $\gamma^\nu$ and
obtain
\begin{equation}
\gamma^\mu[D_\mu,D_\nu]=\ft12[R_{\mu\nu}+4(d-1)m^2g_{\mu\nu}]\gamma^\nu.
\end{equation}
This leads directly to the Einstein condition
\begin{equation}
R_{\mu\nu}=-4(d-1)m^2g_{\mu\nu},
\end{equation}
so we see that the resulting space must be Einstein and have vanishing Weyl holonomy
\begin{equation}
[D_\mu,D_\nu]\epsilon=\ft14C_{\mu\nu\rho\sigma}\gamma^{\rho\sigma}\epsilon.
\end{equation}

Although these spaces are often interesting in their own right (such as manifolds with G$_2$
structure), we instead wish to focus on a different situation, where the deformation of the
space may be compensated for by a modification of the Killing spinor equation.  In particular,
we are interested in preserving a Killing spinor while turning on a background gauge field
$A_\mu$.  Under the appropriate circumstances in the context of the AdS/CFT correspondence, the time component of this gauge field will
admit an interpretation as a chemical potential. This interpretation constitutes an important motivation for us as it could provide a useful generalization of the results reported in \cite{Hama:2011ea} and  \cite{Imamura:2011wg}.

By turning on a background gauge field and considering a charged Killing spinor, we modify
the supercovariant derivative (\ref{eq:Dorig}) into
\begin{equation}
D_\mu=\nabla_\mu+m\gamma_\mu+iA_\mu+\gamma_\mu{}^\nu B_\nu.
\end{equation}
Here we have allowed a second background vector field $B_\mu$ compatible with the
Lorentz and Dirac structure of the supercovariant derivative.  Integrability then
gives
\begin{eqnarray}
[D_\mu,D_\nu]&=&iF_{\mu\nu}+(2\partial_{[\mu}m\delta_{\nu]}^\lambda
-4mB_{[\mu}\delta_{\nu]}^\lambda)\gamma_\lambda\nonumber\\
&&+(\ft14R_{\mu\nu}{}^{\lambda\sigma}
+2(m^2-B^2)\delta_\mu^\lambda\delta_\nu^\sigma-2\delta_{[\nu}^\sigma\nabla_{\mu]}B^\lambda
+4\delta_{[\nu}^\sigma B_{\mu]}B^\lambda)\gamma_{\lambda\sigma}.
\label{eq:abint}
\end{eqnarray}

For completely unbroken supersymmetry, each of the quantities multiplying the different
Dirac matrix combinations in (\ref{eq:abint}) must vanish independently.  This gives rise
to the following conditions for completely unbroken supersymmetry:
\begin{eqnarray}
&&F_{\mu\nu}=0,\qquad G_{\mu\nu}=0,\qquad (\partial_\mu-2B_\mu)m=0,\nonumber\\
&&R_{\mu\nu}=-4(d-1)(m^2-B^2)g_{\mu\nu}-4(B^2g_{\mu\nn}+(d-2)B_\mu B_\nu)
\nonumber\\
&&\kern4em+2(\nabla\cdot Bg_{\mu\nu}+(d-2)\nabla_{(\mu}B_{\nu)}),\nonumber\\
&&C_{\mu\nu\lambda\sigma}=0,
\end{eqnarray}
where $F=dA$ and $G=dB$.  Note that if we take $m$ to be a non-zero constant, then the
condition $(d-2B)m=0$ requires $B$ to vanish.  The system then simplifies to
\begin{equation}
F_{\mu\nu}=0,\qquad
R_{\mu\nu}=-4(d-1)m^2g_{\mu\nu},\qquad
C_{\mu\nu\lambda\sigma}=0.
\label{eq:mconst}
\end{equation}

In general, all we really demand is at least one unbroken supersymmetry.  In this case,
the above conditions are in general too restrictive, and we ought to examine the
integrability expression (\ref{eq:abint}) in its entirety.  It nevertheless seems reasonable
to take $m$ to be a constant and to set $B=0$.  In this case
\begin{equation}
D_\mu=\nabla_\mu+m\gamma_\mu+iA_\mu,
\label{eq:modkse}
\end{equation}
and
\begin{equation}
[D_\mu,D_\nu]=iF_{\mu\nu}+\ft14[R_{\mu\nu}{}^{\lambda\sigma}
+4m^2(\delta_\mu^\lambda\delta_\nu^\sigma-\delta_\mu^\sigma\delta_\nu^\lambda)]
\gamma_{\lambda\sigma}.
\end{equation}
Multiplying this by $\gamma^\nu$ from the left gives
\begin{equation}
\gamma^\nu[D_\mu,D_\nu]=\ft12[R_{\mu\nu}+4(d-1)m^2g_{\mu\nu}+2iF_{\mu\nu}]\gamma^\nu.
\label{eq:asint}
\end{equation}
If we were to split this into symmetric and antisymmetric combinations, we would reproduce
the first two conditions of (\ref{eq:mconst}).  However, more general solutions are allowed
where the Killing spinor satisfies a projection of the form
\begin{equation}
F_{\mu\nu}\gamma^\nu=i\Lambda_{\mu\nu}\gamma^\nu,
\label{eq:Lproj}
\end{equation}
where $\Lambda_{\mu\nu}$ is now a {\it symmetric} matrix.  Substituting this into
(\ref{eq:asint}) then gives the modified Einstein equation
\begin{equation}
R_{\mu\nu}=-4(d-1)m^2g_{\mu\nu}+2\Lambda_{\mu\nu}.
\label{eq:modeins}
\end{equation}

As an example, consider a basis where
\begin{equation}
F_{\mu\nu}=\begin{pmatrix}&\lambda_1\cr-\lambda_1\cr&&&\lambda_2\cr
&&-\lambda_2\cr&&&&\ddots\end{pmatrix}
\end{equation}
It is then straightforward to see that
\begin{equation}
\Lambda_{\mu\nu}=\begin{pmatrix}\eta_1\lambda_1\cr&\eta_1\lambda_1\cr
&&\eta_2\lambda_2\cr&&&\eta_2\lambda_2\cr&&&&\ddots\end{pmatrix},
\end{equation}
provided the Killing spinor $\epsilon$ satisfies the projections
\begin{equation}
\gamma^{12}\epsilon=i\eta_1\epsilon,\qquad
\gamma^{34}\epsilon=i\eta_2\epsilon,\qquad \ldots.
\label{eq:bpsprojs}
\end{equation}
In this case, the deformed background only preserves a fraction of the original supersymmetries,
as determined by the number of independent projections in (\ref{eq:bpsprojs}). This is the case of
the squashed $S^3$ discussed in \cite{Hama:2011ea}.  Furthermore, as we will show in
Appendix~\ref{app:odd}, this also extends to any odd-dimensional sphere.

\section{The embedding construction of the squashed $\mathrm{S}^4$}
\label{app:S4}

Let us consider the squashed $\mathrm{S}^4$ as a solution to the supergravity variations. We deviate from our main discussion in the text by starting with the following Ansatz for the metric:
\be
ds^2 =e^{2\Omega(\theta)}\left(f^2(\theta) d\theta^2 +\sin^2\theta d\Omega_3^2\right).
\ee
A natural generalization of the previous description of the round $\mathrm{S}^4$ can be achieved by introducing the  following squashing in the vierbein
\begin{align}
    e^i = e^{\Omega(\theta)}\sin\theta\bar e^i,
    &&
    e^4 = e^{\Omega(\theta)}f(\theta)d\theta.
\end{align}

Motivated by the $S^3$ squashing discussed in \cite{Hama:2011ea}, it is tempting to take $f(\theta) = \sqrt{\cos^2\theta+\delta^2\sin^2\theta}$, however we let $f$ be an arbitrary function at this point. When $\delta=1$, $f=1$ and $\Omega=0$, we return to the round sphere. The spin connection for this case is
\begin{align}
    \w_{ij}=\bar\w_{ij},
    &&
    \w_{i4}=\frac{\left(\Omega'\sin\theta+\cos\theta\right)}{f}\bar e^i.
\end{align}
Using the map between 3-dim and 4-dim gamma matrices
\begin{align}
    \gamma^i = \sigma^1\otimes\sigma^i,
&&
    \gamma^4 = \sigma^2\otimes\idn,
&&
    \gamma^5 = \sigma^3\otimes\idn,
&
\end{align}
we write the Killing spinor equation as\footnote{$N_1=-\frac12(M+\bar{M})$ and $N_2=\frac1{2i}(\bar{M}-M)$.}
\begin{subequations}
\begin{align}
    0=&\left[\pd_\theta-\frac16e^\Omega f\gamma^4(N_1+i\gamma^5 N_2)
    +i\gamma^5\left(\frac{1}{3}V_\theta-\frac{1}{6}e_\theta^a\gamma_a\gamma_\mu V^\mu\right)\right]\eps,
\\
    0=&\Big[d_{(3)}+\frac14\bar\w_{ij}\gamma^{ij}+\frac12\frac{\left(\Omega'\sin\theta+\cos\theta\right)}{f}\bar e^i\gamma^{i4}
    +\cr&
    -\frac16e^\Omega\sin\theta \bar e^i\gamma^i(N_1+i\gamma^5 N_2)
    +i\gamma^5\left(\frac{1}{3}V_{(3)}-\frac{1}{2}e^\Omega \sin\theta \bar e^i\gamma^i\gamma_\mu V^\mu\right)\Big]\eps.
\end{align}
\end{subequations}
Note that in the second equation we have used form notation. We write the spinor using a 3-dim Killing spinor (round sphere) and unknown 2 component vector c ($\theta$ dependant)
\begin{align}
    \eps = c(\theta)\otimes\eta,
    &&
    \left(d_{(3)}+\frac14\bar\w_{ij}\sigma^{ij}\right)\eta = \frac{i}2\bar e^i\s^i \eta.
\end{align}
We also make the following assumptions,
\begin{align}
    N_1 = 3i\frac{M(\theta)}{e^{\Omega(\theta)}f(\theta)}\cos\alpha,
&&
    N_2 = 3i\frac{M(\theta)}{e^{\Omega(\theta)}f(\theta)}\sin\alpha,
&&&
    V = -3i\,b_4(\theta)e^4.
\end{align}
Hence,
\begin{equation*}
    \gamma_\mu V^\mu = -3i\frac{b_4}{e^\Omega f}\gamma^4.
\end{equation*}

We now use the decomposition of the 4-dim gamma matrices in terms of the 3-dim sigma matrices to arrive at
\begin{subequations}
\begin{align}
    0=&\left[\pd_\theta
    -\frac i2M \s^2(\cos\alpha+i\s^3\sin\alpha)
    +\frac{1}{2}\s^3b_4\right]\chi(\theta)\otimes\eta,
    \label{eq:eqn1}
\\
    0=&\Big[\left(1-b_4\frac{\sin\theta}{f}\right)\idn
    +\frac{\left(\Omega'\sin\theta+\cos\theta\right)}{f}\s^3
    +\cr&
    -\frac{M}{f}\sin\theta\, \s^1(\cos\alpha+i\s^3\sin\alpha)
    \Big]\chi(\theta)\otimes\left(\frac i2 \bar e^i\s^i\eta\right).
    \label{eq:eqn2}
\end{align}
\end{subequations}
We proceed by making sure that the second equation (the non-differential one) has a non trivial solution. First we introduce new functions
\begin{align*}
    B = f-b_4\sin\theta,
    &&
    C=\Omega'\sin\theta+\cos\theta.
\end{align*}
The determinant of (\ref{eq:eqn2}) now takes the form
\begin{equation}\label{sqh-cnstr}
    B^2-C^2-M^2\sin^2\theta =0.
\end{equation}
We may also directly solve (\ref{eq:eqn2}) for $\chi_1$ in terms of $\chi_2$ and insert this
into (\ref{eq:eqn1}).
After some manipulation, we may obtain a second constraint
\begin{equation}
    \left(\frac{M\sin\theta }{B+C}\right)'
    +\frac{M\sin\theta}{B+C}b_4-\frac12M
    =
    \frac{M^3\sin^2\theta}{2(B+C)^2}.
\end{equation}
Using (\ref{sqh-cnstr}), we can simplify the above and summarize the two constraints that we need to solve\footnote{It is important to note that the definition of $\theta$ guarantees that $\sin\theta\geq0$.},
\begin{subequations}
\begin{align}\label{sqh-constrains}
    M^2\sin^2\theta =& B^2-C^2,
\\
    \frac{1}{2}f
    =&
    B
    -\frac{1}{2}\frac{B'C-BC'}{B^2-C^2}\sin\theta.
\end{align}
\end{subequations}
Let us consider a simple particular solution. Choose $\Omega=0$ and $M=\d$ (a constant). We easily find
\begin{align*}
    B^2=&\cos^2\theta+\d^2\sin^2\theta,
\\
    f
    =& 2\frac{\cos^2\theta+\d^2\sin^2\theta-1/2}{\sqrt{\cos^2\theta+\d^2\sin^2\theta}},
\\
    b_4 =& -\frac{(1-\d^2)\sin\theta}{\sqrt{\cos^2\theta+\d^2\sin^2\theta}}.
\end{align*}
We conclude with an analysis of the regularity of the solution. For $f > 0$ we need $\d^2>1/2$. Then we need to check what happens near the north and south poles of the squashed sphere
\begin{align*}
    f(\theta) \simeq & 1 + O(\theta^2),
    &
    b(\theta) \simeq & -(1-\d^2)\theta + O(\theta^3),
\\
    f(\theta) \simeq & 1 + O((\pi-\theta)^2),
    &
    b(\theta) \simeq & -(1-\d^2)(\pi-\theta) + O((\pi-\theta)^3).
\end{align*}
Therefore, the solution is regular around the only two points where potential singularities might occur.

Let us summarize the role of squashing from the geometric point of view. It is not hard to see that $\Omega$ and $f$ are a combination of a diffeomorphism and a Weyl transformation. Since we started with a space with vanishing Weyl tensor, the squashing we introduce keeps the Weyl tensor zero. It is interesting  that the vector $b$ is closed and exact
\begin{equation}
    V_{(1)} = -3ib_4e^\Omega f d\theta
    =6i(1-\d^{-2})\,d\left[\d^2\cos\theta-\frac{1}{2\sqrt{1-\d^{-2}}}\;\mathrm{arctanh}\,\left(\sqrt{1-\d^{-2}}\cos\theta\right)\right].
\end{equation}
It is also curious that $V_{(1)}$ is imaginary. We know that for $N_i$ this is okay, and unitarity will not be violated. However it is not clear this is true for $V_{(1)}$.

\section{Killing spinors and squashing spheres in various dimensions}\label{app:odd}
\subsection{Odd dimensional squashed spheres}

The introduction of a vector field in the Killing spinor equation is natural for spaces admitting
a U(1) fibration.  As an example, we consider the Hopf fibration of $S^{2n+1}$ over $\mathbb{CP}^n$.
To set up the analysis, consider first the round unit $S^{2n+1}$, with metric given by
\begin{equation}
d\Omega_{2n+1}^2=ds^2(\mathbb{CP}^n)+(d\psi+\mathcal A)^2,
\end{equation}
where $d\mathcal A=2J$ and $J$ is the Kahler form on $\mathbb{CP}^n$.  The round unit sphere admits
a complete set of Killing spinors $\epsilon$ satisfying
\begin{equation}
\hat D_A\epsilon\equiv(\nabla_A+\ft{i}2\gamma_A)\epsilon=0.
\end{equation}
This corresponds to taking $m=i/2$ in (\ref{eq:Dorig}).
Using a natural vielbein basis
\begin{equation}
d\Omega_{2n+1}^2=\hat e^a\otimes\hat e^a+\eta\otimes\eta,
\end{equation}
the Killing spinor equation decomposes as
\begin{eqnarray}
\hat D_a&=&\hat\nabla_a-\mathcal A_a\partial_\psi+\ft12J_{ab}\gamma^{b9}+\ft{i}2\gamma_a,
\nonumber\\
\hat D_9&=&\partial_\psi-\ft14J_{ab}\gamma^{ab}+\ft{i}2\gamma^9,
\end{eqnarray}
where $\hat e^9\equiv\eta=d\psi+\mathcal A$.

In this decomposition, the Killing spinors are charged along the U(1) fiber.  Taking
$\partial_\psi\to iq$, we see that $\hat D_9\epsilon=0$ requires the charge condition
\begin{equation}
q=-\ft12(\ft{i}2J_{ab}\gamma^{ab}+\gamma^9)
=-\ft12(i\gamma^{12}+i\gamma^{34}+\cdots+\gamma^9),
\label{eq:ccond}
\end{equation}
where each of the Dirac matrix factors in the last expression has eigenvalues $\pm1$.
For $\hat D_a\epsilon=0$, we examine the integrability condition
\begin{equation}
[\hat D_a,\hat D_b]=\ft14[\hat R_{ab}{}^{cd}-J_a{}^cJ_b{}^d+J_a{}^dJ_b{}^c-2J_{ab}J^{cd}
-(\delta_a^c\delta_b^d-\delta_a^d\delta_b^c)]\gamma_{cd}
-iJ_{ab}(2q+\ft{i}2J_{cd}\gamma^{cd}+\gamma^9).
\end{equation}
The integrability condition vanishes identically for Killing spinors, as the first term corresponds to the Riemann
tensor on $\mathbb{CP}^n$, and the second term is identical to the charge condition (\ref{eq:ccond}).

We now turn to the squashed sphere, with metric
\begin{equation}
ds^2=\ell^2ds^2(\mathbb{CP}^n)+\tilde\ell^2(d\psi+\mathcal A)^2,
\end{equation}
and ask whether it admits Killing spinors satisfying the modified Killing spinor equation
(\ref{eq:modkse}).  We first compute the curvature:
\begin{eqnarray}
\ell^2R_{ab}{}^{cd}&=&\hat R_{ab}{}^{cd}-\alpha^2(J_a{}^cJ_b{}^d
-J_a{}^dJ_b{}^c+2J_{ab}J^{cd}),\nonumber\\
\ell^2R_{a9}{}^{b9}&=&\alpha^2\delta_a^b,
\end{eqnarray}
and the Ricci tensor
\begin{equation}
\ell^2R_{ab}=\hat R_{ab}-2\alpha^2\delta_{ab},\qquad
\ell^2R_{99}=(d-1)\alpha^2,
\end{equation}
where $\alpha=\tilde\ell/\ell$ is the squashing parameter.  Note that tangent space components
of $\hat R$ and $J$ are obtained using the $\hat e^a$ vielbeins while those of $R$ are obtained
using  $E^a$ where $E^a=\ell\hat e^a$ and $E^9=\tilde\ell\eta$.

Since we take the curvature of $\mathbb{CP}^n$ to be $\hat R_{ab}=(d+1)\delta_{ab}$, the base of
the fibration satisfies the Einstein condition
\begin{equation}
\ell^2R_{ab}=(d+1-2\alpha^2)\delta_{ab}.
\end{equation}
As a result, the modified Einstein equation (\ref{eq:modeins}) takes the form
\begin{eqnarray}
2\ell^2\Lambda_{ab}&=&-(d+1-2\alpha^2+(d-1)(2m\ell)^2)\delta_{ab},\nonumber\\
2\ell^2\Lambda_{99}&=&-(d-1)(\alpha^2+(2m\ell)^2),
\label{eq:einsL}
\end{eqnarray}
where $\Lambda_{AB}$ is associated with the background vector according to the projection
(\ref{eq:Lproj}):
\begin{equation}
F_{AB}\gamma^B=i\Lambda_{AB}\gamma^B.
\end{equation}
Since there is only one natural 2-form on $\mathbb{CP}^n$, namely $J$, we anticipate taking
$F_{ab}$ proportional to $J_{ab}$ along with $\Lambda_{ab}$ proportional to $\delta_{ab}$.
In particular, we set $\Lambda_{99}=0$, in which case (\ref{eq:einsL}) may be solved by
taking $m\ell=i\alpha/2$ and
\begin{equation}
\ell^2\Lambda_{ab}=\ft12(d+1)(\alpha^2-1)\delta_{ab}.
\end{equation}
This is compatible with the field strength
\begin{equation}
F=\ft12(d+1)(\alpha^2-1)J,
\end{equation}
or equivalently $A=\fft14(d+1)(\alpha^2-1)\mathcal A$.

To complete the discussion we need to check not only the integrability condition but also the actual Killing spinor equation. The latter analysis shows that the background preserves $1/2^n$ supersymmetries.  In particular, the solution of \cite{Hama:2011ea} corresponding to the squashed $S^3$ is explicitly half-supersymmetric.

What this indicates is that the squashed $S^{2n+1}$ admits a Killing spinor satisfying
$D_\mu\epsilon=0$ where
\begin{equation}
D_\mu=\nabla_\mu+\fft{i\alpha}{2\ell}\gamma_\mu+\fft{i(d+1)}4(\alpha^2-1)\mathcal A_\mu,
\end{equation}
and where $\alpha=\tilde\ell/\ell$.

\subsection{Distorted spheres in arbitrary dimensions}

While odd-dimensional spheres admit a natural squashing along a U(1) fiber,
even-dimensional spheres cannot be treated in the same manner.  There is, however, an
alternate method to distorting the sphere.  Consider, for example, a round $n$-sphere
embedded in $(n+1)$-dimensional Euclidean space.  The metric on the sphere is then
the Euclidean metric
\begin{equation}
ds^2=\delta_{ij}dx^idx^j,
\label{eq:Emet}
\end{equation}
restricted to the surface of the sphere, $\delta_{ij}x^ix^j=R^2$.  We now distort this round
sphere into an ellipsoid by taking
\begin{equation}
T_{ij}x^ix^j=1,
\end{equation}
where $T_{ij}$ is a constant symmetric matrix.  The normal to the distorted sphere is given by
\begin{equation}
n_i=\fft{T_{ij}x^j}{\sqrt\Delta},\qquad\mbox{where}\qquad\Delta=\vec x\,T^2\vec x=
x^iT_{ij}T_{jk}x^k,
\end{equation}
and the induced metric is simply $h_{ij}=\delta_{ij}-n_in_j$.

Before determining the Killing spinors on the distorted sphere, we first examine some
general properties.  Defining the projection of $T_{ij}$ onto the sphere by
\begin{equation}
\tilde T_{ij}=h_{ik}T_{kl}h_{lj},
\end{equation}
we find the extrinsic curvature tensor to be
\begin{equation}
K_{ij}=\fft{\tilde T_{ij}}{\sqrt\Delta},
\end{equation}
in which case the Gauss-Codazzi equations give
\begin{equation}
\tilde R_{ijkl}=\Delta^{-1}(\tilde T_{ik}\tilde T_{jl}-\tilde T_{il}\tilde T_{jk}).
\end{equation}
The condition for maximal symmetry, (\ref{eq:Rmax}), is then equivalent to
\begin{equation}
\tilde T_{ij}=\sqrt{-4m^2\Delta}\, h_{ij}.
\end{equation}
This is solved for a round sphere by taking $T_{ij}=\delta_{ij}/\ell^2$, so that $\sum (x^i)^2=\ell^2$
and $\Delta=1/\ell^2$.  In this case, we obtain the familiar result $m^2=-1/(2\ell)^2$.

A general ellipsoid may be written in terms of a set of principle axes by taking
\begin{equation}
T_{ij}=\fft{\delta_{ij}}{\ell_i^2}\qquad(\mbox{no sum}).
\end{equation}
However, we are primarily interested in an axisymmetric distortion, so we take
\begin{equation}
T_{ij}=\mbox{diag}(\ell^{-2},\ell^{-2},\tilde\ell^{-2},\ldots,\tilde\ell^{-2}).
\end{equation}
For this parameterization, we find it more convenient to work in an unconstrained basis by
defining
\begin{equation}
x^1=\ell\cos\theta\cos\phi,\qquad
x^2=\ell\cos\theta\sin\phi,\qquad
x^a=\tilde\ell\sin\theta\mu^a,
\end{equation}
where $a=3,4,\ldots,n+1$ and $\sum_a(\mu^a)^2=1$ defines the unit $S^{n-2}$.  The
metric on the sphere, (\ref{eq:Emet}), is then given by
\begin{equation}
ds^2=f\,d\theta^2+\ell^2\cos^2\theta\,d\phi^2+\tilde\ell^2\sin^2\theta\,d\Omega_{n-2}^2,
\label{eq:distSn}
\end{equation}
where
\begin{equation}
f=\sqrt{\ell^2\sin^2\theta+\tilde\ell^2\cos^2\theta}.
\end{equation}

Since we are looking for Killing spinors on the distorted sphere, we begin with the unmodified
covariant derivative (\ref{eq:Dorig}).  Working out the spin connections in the obvious vielbein
basis
\begin{equation}
e^1=\sqrt{f}\,d\theta,\qquad e^2=\ell\cos\theta\,d\phi,\qquad e^a=\tilde\ell\sin\theta\,\hat e^a,
\label{eq:Snviels}
\end{equation}
we find
\begin{eqnarray}
D_\theta&=&\partial_\theta+mf\gamma^1,\nonumber\\
D_\phi&=&\partial_\phi+\ft12\ell f^{-1}\sin\theta\gamma^{12}+m\ell\cos\theta\gamma^2,
\nonumber\\
D_\alpha&=&\hat\nabla_\alpha-\ft12\tilde\ell f^{-1}\hat e_\alpha^a\gamma^{1a}
+m\tilde\ell\sin\theta\hat e_\alpha^a\gamma^a.
\label{eq:oldDdist}
\end{eqnarray}
For the round sphere, we have $\ell=\tilde\ell$, so that $f=1$ and $m=i/2$.  However, for $f\ne1$,
it is clear that the original Killing spinor $\epsilon$ will no longer satisfy $D_\mu\epsilon=0$
because of the factors of $f$ showing up in (\ref{eq:oldDdist}).

Of course, the explicit terms in (\ref{eq:oldDdist}) gives us a hint as to what must be done to
modify the supercovariant derivative.  Starting with $D_\theta$, we see that $m$ should now
be set to $i/2f$.  Inserting this into $D_\phi$ then gives
\begin{equation}
D_\phi=\partial_\phi+\ft12\ell f^{-1}(\sin\theta\gamma^{12}+i\cos\theta\gamma^2).
\end{equation}
In order to compensate for the $\ell/f$ factor, we may introduce a gauge field,
\begin{equation}
A=\fft12\left(1-\fft\ell{f}\right)d\phi
\end{equation}
which only modifies $D_\phi$.  Finally, turning to $D_\alpha$, we see a similar need to
compensate for the $\tilde\ell/f$ factor.  However, since this applies to all directions on
$S^{n-2}$, what is needed is something of the form
\begin{equation}
A=\fft12\left(1-\fft{\tilde\ell}f\right)\gamma^a\hat e^a.
\end{equation}
The problem with such an expression is that it has an unwanted Dirac matrix $\gamma^a$,
and as a result does not admit a covariant extension to the full $S^n$.  One way to get
around this problem is to add a background antisymmetric tensor $C_{\mu\nu}$ to the
supercovariant derivative, so that we end up with
\begin{equation}
D_\mu=\nabla_\mu+m\gamma_\mu+iA_\mu+\ft12\gamma_\mu{}^{\nu\rho}C_{\nu\rho}.
\label{eq:newD}
\end{equation}
We may then set
\begin{equation}
C=\fft{f\ell\cos\theta}{2\tilde\ell\sin\theta}\left(1-\fft{\tilde\ell}f\right)d\theta\wedge d\phi.
\end{equation}
In particular, the Dirac structure in (\ref{eq:newD}) is designed so that $C$ will only contribute
to the components of $D_\mu$ along $S^{n-2}$.

To summarize, we have found Killing spinors on the distorted sphere (\ref{eq:distSn})
using a modified supercovariant derivative (\ref{eq:newD}).  The background fields are
\begin{eqnarray}
m&=&\fft{i}{2f},\nonumber\\
A&=&\fft1{2\ell\cos\theta}\left(1-\fft\ell{f}\right)e^2,\nonumber\\
C&=&\fft1{2\tilde\ell\sin\theta}\left(1-\fft{\tilde\ell}f\right)e^1\wedge e^2,
\end{eqnarray}
where the vielbeins are given in (\ref{eq:Snviels}).  The field strength of $A$ is given by
\begin{equation}
F=dA=\fft{\sin\theta}{2f^4}\left(\ell^2-\tilde\ell^2\right)e^1\wedge e^2,
\end{equation}
so both $F$ and $C$ are only turned on in the directions orthogonal to $S^{n-2}$.  Of course,
both of these fields vanish in the round sphere limit, as expected.  Furthermore,
unlike for the Hopf fibration example, $m$ is not a constant in this case.

\subsection{Killing spinors of squashed $S^{n+1}$ --- the embedding approach}

We study possible Killing spinors for the metric coming from squashing the embedding space. We break the embedding space coordinate into two parts,
\begin{equation}
    \frac1{l^2}\sum_{i=1}^{n+1}(X^i)^2+\frac1{\tilde l^2}\sum_{i=1}^{m+1}(Y^i)^2 = 1.
\end{equation}
We write the background metric with a conformal factor
\begin{equation}\label{sqm}
    ds^2 = e^{2\Omega}\left(fd\theta^2+l^2\cos^2\theta d\Omega_{m}^2+\tilde l^2\sin^2\theta d\Omega_{n-m}^2\right),
\end{equation}
with $f=\sqrt{l^2\sin^2\theta+\tilde l^2\cos^2\theta}$, and $\Omega$ a function of $\theta$ to be determined later. In the round sphere limit $\tilde l=l$, $f=l$ and $\Omega=0$.
We add to the background gravity a mass $M$, a vector $A$ and a field strength $F_{(m+1)}=dC_{(m)}$, where the $m$-form potential $C_{(m)}$ is in the $S^{m}$ directions. The Killing spinor equation we consider takes the form
\begin{multline}
    d\eps+\frac14w_{ab}\gamma^{ab}\eps+iA\eps+iM\gamma^ae^a\eps+\frac{\tilde\alpha}{(m)!}F_{A_1A_2\ldots A_{m+1}}\gamma^{A_1A_2\ldots A_{m}}\eps
    +\\
    +\frac{\tilde\beta}{(m+1)!}F_{A_1A_2\ldots A_{m+1}}\gamma^{A_1A_2\ldots A_{m+1}}_{A_{m+2}}\eps=0,
\end{multline}
with $\tilde\alpha=i^{\left[\frac{m+2}2\right]}\sqrt{\frac{n-m-1}{8m(n-1)}}$ and $\tilde\beta=-\frac{m}{n-m-1}\tilde\alpha$. The capital Latin letter indices are coordinate indices, while the small Latin letter indices are local frame indices. We are interested in partial supersymmetry, and do not look at the integrability condition in this case.

We choose $C_{(m)}$ to be proportional to the volume form on $S^{m}$, up to a function of $\theta$. Thus the field strength is
\begin{equation}
    F_{(m+1)}=p'(\theta)e^1\wedge e^2\cdots \wedge e^m\wedge e^{n+1}
\end{equation}
Setting the notation, indices on the squashed $S^{n+1}$ will be given by $a,b=1,2\ldots n+1$,  indices in the $S^m$ direction by $i,j=1,2\ldots m$ and indices in the $S^{n-m}$ direction by $\tilde i,\tilde j=m+1,m+2\ldots n$. Quantities with a bar come from the round $S^{m}\times S^{n-m}$ space, where both spheres are unit spheres.
The non-coordinate basis we use is
\begin{align}
&    e^{i}=e^{\Omega}l\cos\theta \bar e^i, &
&    e^{\tilde i}=e^{\Omega}\tilde l\sin\theta \bar e^{\tilde i}, &
&    e^{n+1}=e^{\theta}=e^{\Omega}fd\theta.&
\end{align}
The spin connection is
\begin{subequations}
\begin{align}
    \o_{i\theta}=&\,-\o_{\theta i}=\left(\Omega'-\tan\theta\right)\frac{l\cos\theta}{f}\bar e^i,
\\
    \o_{i\theta}=&\,-\o_{\theta i}=\left(\Omega'+\cot\theta\right)\frac{l\sin\theta}{f}\bar e^i,
\\
    \o_{ij}=&\,\bar \o_{ij},
\\
    \o_{\tilde i\tilde j}=&\,\bar \o_{\tilde i\tilde j},
\\
    \o_{i\tilde i}=&\,0.
\end{align}
\end{subequations}
Putting it all together, we can write the Killing spinor equation as
\begin{subequations}
\begin{align}
\label{kse1}
    \bar d_{(m)}\eps+\frac14\bar w_{ij}\gamma^{ij}\eps
    -\frac12\left(\Omega'-\tan\theta\right)\frac{l\cos\theta}{f}\gamma^\theta\gamma^{i}\bar e^i\eps
    +iA_{(m)}\eps
    +&\cr
    +iMe^\Omega l\cos\theta\gamma^i\bar e^i\eps
    +\alpha e^{\Omega}p'\,l\cos\theta\Gamma^{(m)}\gamma^{\theta}\gamma^i\bar e^i\eps &=0,
\\
\label{kse2}
    \bar d_{(n-m)}\eps+\frac14\bar w_{\tilde i\tilde j}\gamma^{\tilde i\tilde j}\eps
    -\frac12\left(\Omega'+\cot\theta\right)\frac{\tilde l\sin\theta}{f}\gamma^\theta\gamma^{\tilde i}\bar e^{\tilde i}\eps
    +iA_{(n-m)}\eps    +&\cr
    +iMe^\Omega \tilde l\sin\theta\gamma^{\tilde i}\bar e^{\tilde i}\eps
    +\beta e^\Omega p'\,\tilde l\sin\theta\Gamma^{(m)}\gamma^\theta\gamma^{\tilde i}\bar e^{\tilde i}\eps &=0,
\\
\label{kse3}
    d_\theta\eps+iA_\theta\eps+iMe^\Omega f\gamma^\theta\eps+\alpha e^{\Omega}p'f\,\Gamma^{(m)}\eps&=0,
\end{align}
\end{subequations}
with $\Gamma^{(m)}=\gamma^{i_1i_2\ldots i_{m}}$. Notice that $\alpha$ and $\beta$ still need to be calculated (combinatorically).

One approach to finding the background for the squashed sphere is to find $p$ and $\Omega$ such that the Killing spinor equation is reduced to the round sphere case. This approach guarantees the smoothness of the limit $\tilde l\rightarrow \l$. Taking this (limiting) approach, we can examine the round sphere limit of \eqref{kse3}. We learn that $\eps$ must be an eigenvector of $\gamma^\theta$, and this should carry over to the squashed case too. From the round sphere case we also can see that we need $\eps$ to be an eigenvector of $\Gamma^{(m)}$. This can be done only if $\gamma^\theta$ and $\Gamma^{(m)}$ commute, which forces $m$ to be even (including $m=0$).

In the following we assume that $m$ is even and $\eps$ is indeed an eigenvector of $\gamma^\theta$ and $\Gamma^{(m)}$, so we can interpret these matrices as $c$-numbers. More precisely, $\gamma^\theta=\pm1$ and $\Gamma^{(m)}=\pm i$. We can now find the background field that `eliminates' the effect of the squashing
\begin{subequations}
\begin{align}
    A_{(m)} = A_{(n-m)} =&0,
\\
    \frac{l\Omega'\cos\theta}{2f}
    -\alpha e^{\Omega}p'\,l\cos\theta\Gamma^{(m)}
    =&\,
    \left(\frac{l}{f}
    -1\right)\frac{\sin\theta}{2}
    -i\left(Me^\Omega-m\right)l\cos\theta\,\gamma^\theta,
\\
    \frac{\tilde l\Omega'\sin\theta}{2f}
    -\beta e^\Omega p'\,\tilde l\sin\theta\Gamma^{(m)}
    =&\,
    -\left(\frac{\tilde l}{f}
    -1\right)\frac{\cos\theta}2
    -i\left(Me^\Omega-m\right)l\sin\theta\,\gamma^\theta,
\\
    A_\theta=&\,-\left(\frac{Mfe^\Omega}{ml}-1\right)ml\gamma^\theta+i\alpha e^{\Omega}p'f\,\Gamma^{(m)},
\end{align}
\end{subequations}
where $m$ is the mass parameter in the round sphere case. The value of $M$ compared to $m$ should come from the CFT coupling and is not a background parameter. The solution to the above set of equations has potential singularities near $\theta =0$ and $\theta=\frac{\pi}{2}$; we can verify that the solution is smooth by considering
\begin{align}
    &\left(\frac{l}{f}-1\right)\sin\theta\xrightarrow{\theta\rightarrow\frac{\pi}2}\frac{l^2-\tilde l^2}{2l^2}\theta^2,&
    &\left(\frac{\tilde l}{f}-1\right)\cos\theta\xrightarrow{\theta\rightarrow0}\frac{\tilde l^2-l^2}{2\tilde l^2}\theta^2.&
\end{align}
However, we still need to break these equations to their real and imaginary parts. Note that $\Omega, M$ are real and $\Gamma^{(m)}$ is imaginary. Then if $\alpha p'\Gamma^{(m)}$ is real  we get a complex $A_\theta$. If this combination is imaginary we find two different differential equations for $\Omega(\theta)$ and thus have no solution.  Assuming we can have a complex $A$, then we still need to have $M = e^{-\Omega} m$.  This is not a background field choice but a specific supergravity coupling.

\bibliographystyle{JHEP}
\bibliography{SusyFT-bib}

\end{document}